\documentclass{JHEP3}
\usepackage[dvips]{graphicx}
\usepackage{latexsym}
\usepackage{epsfig}

\newcommand{\lya}{Lyman-$\alpha$~}
\newcommand{\smnu}{$\Sigma m_{\nu}$~}
\def\b{\begin{equation}}
\def\ee{\end{equation}}
\def\bea{\begin{eqnarray}}
\def\eea{\end{eqnarray}}

\title{The effect of neutrinos on the matter distribution as probed by
  the Intergalactic Medium} \author{Matteo Viel${}^{1,2}$, Martin
  G. Haehnelt${}^{3,4}$, Volker Springel${}^5$
  \\ ${}^1$INAF-Osservatorio Astronomico di Trieste, Via G.B. Tiepolo
  11, I-34131 Trieste, Italy \\ ${}^2$INFN sez. Trieste, Via Valerio
  2, 34127 Trieste, Italy\\ ${}^3$Institute of Astronomy, Madingley
  Road, CB3 0HA, UK\\ ${}^4$KICC-Kavli Insitute of Cosmology,
  Cambridge, UK\\ ${}^5$Max-Planck Institut f\"ur Astrophysik,
  Karl-Schwarzschild Str. 1, 85748 Garching, Germany\\ Email:
  \email{viel@oats.inaf.it},\email{haehnelt@ast.cam.ac.uk},\email{volker@mpa-garching.mpg.de}}

\abstract{We present a suite of full hydrodynamical cosmological
  simulations that quantitatively address the impact of neutrinos on
  the (mildly non-linear) spatial distribution of matter and in
  particular on the neutral hydrogen distribution in the Intergalactic
  Medium (IGM), which is responsible for the intervening \lya
  absorption in quasar spectra. The free-streaming of neutrinos
  results in a (non-linear) scale-dependent suppression of power
  spectrum of the total matter distribution at scales probed by
  Lyman-$\alpha$ forest data which is larger than the linear theory
  prediction by about 25~\% and strongly redshift dependent. By
  extracting a set of realistic mock quasar spectra, we quantify the
  effect of neutrinos on the flux probability distribution function
  and flux power spectrum. The differences in the matter power spectra
  translate into a $\sim2.5\%$ ($5\%$) difference in the flux power
  spectrum for neutrino masses with $\Sigma m_{\nu} = 0.3$ eV (0.6
  eV).  This rather small effect is difficult to detect from present
  Lyman-$\alpha$ forest data and nearly perfectly degenerate with the
  overall amplitude of the matter power spectrum as characterised by
  $\sigma_8$. If the results of the numerical simulations are
  normalized to have the same $\sigma_8$ in the initial conditions,
  then neutrinos produce a smaller suppression in the flux power of
  about 3\% (5\%) for $\Sigma m_{\nu} = 0.6$ eV (1.2 eV) when compared to a
  simulation without neutrinos. We present constraints on neutrino
  masses using the Sloan Digital Sky Survey flux power spectrum alone
  and find an upper limit of $\Sigma m_{\nu} < 0.9$ eV (2$\sigma$
  C.L.), comparable to constraints obtained from the cosmic microwave
  background data or other large scale structure probes.}

\begin{document}

\section{Introduction}

One of the most exciting results in particle physics in the last
decade has been that neutrinos have been established to be massive
particles. Solar, atmospheric, reactor and accelerator neutrino
experiments have confirmed the existence of flavour oscillations of
active neutrinos, implying that neutrinos have non-zero mass (see
Ref.~\cite{lespast} and references therein).  This is generally
considered as definite evidence for new physics beyond the Standard
Model.  The neutrino oscillation experiments do, however, not pin down
the absolute neutrino masses. The experiments instead provide a lower
limit for the sum of the neutrino masses of $0.05-0.1$ eV.  Current
measurement of the matter power spectrum from Cosmic Microwave
Background (CMB) data extrapolated to smaller scales alone already
give an upper limit on the sum of the neutrino masses of about $1.5$
eV well below what has been reached with particle physics experiments
leaving an allowed range of only a factor about twenty for the sum of
the neutrino masses. There is thus very strong motivation to push hard
for an actual measurement of neutrino masses.  The tritium
$\beta$-decay experiment
{\small{KATRIN}}\footnote{http://www-ik.fzk.de/$\sim$katrin} is the most
ambitious current direct detection experiment and is expected to probe
an electron neutrino mass of $\sim 0.2$ eV in the near future (see
\cite{fogli} for a recent review).

The matter distribution in the Universe is sensitive to the
free-streaming of cosmological neutrinos. Astrophysical constraints are therefore a
very competitive alternative method to measure/constrain the masses of
neutrinos.  Measurements of the matter power spectrum can in principle
probe neutrino masses significantly smaller than the upper limit from
CMB experiments. Early on the neutrinos are relativistic and travel at
the speed of light with a free-streaming length equal to the Hubble
radius.  Neutrinos in the mass range $0.05$ eV $\le \Sigma m_{\nu} \le
1.5$ eV, become non-relativistic in the redshift range
$3000\ge z \ge 100$.  In the mass range of degenerate neutrino masses 
the thermal velocities can be approximated as,
\begin{equation}
v_{\rm th}\sim 150\, (1+z)\,\left[\frac{1\,\rm{eV}}{\Sigma m_{\nu}}\right] \rm {km/s} \,.
\label{eqvel}
\end{equation}
As a result present-day velocities (of the most massive neutrino
species) range between 100 km/s for the upper and and 3000 km/s for
the lower end of the still allowed range of the sum of the neutrinos
masses. Dark matter particles with such a high velocity dispersion are
usually called hot dark matter. A dominant contribution of hot dark
matter to the total dark matter content would be at odds with current
observations. Neutrinos in the still allowed mass range instead
constitute a sub-dominant contribution complementing cold dark matter
comprised of some other elementary particle, such as neutralinos or
axions.

The effect  of cosmological neutrinos on the evolution of density
perturbation  in the  linear regime is well understood. Neutrinos
affect both the cosmic expansion rate and the growth of structure 
(\cite{maber95,lespast}).  The neutrino contribution in terms of energy density can be expressed as:
\begin{equation}
f_{\nu}=\Omega_{0\nu}/\Omega_{\rm 0m}\,,\,\,\,\,\,\,\,\,\Omega_{0\nu}=\frac{\Sigma\,
  m_{\nu}}{93.8\,h^2\rm{eV}},
\end{equation}
where $h$ is the present value of the Hubble constant in units of 100 km/s/Mpc
and $\Omega_{\rm 0m}$ is the matter energy density in terms of  the critical
density.

When neutrinos become non relativistic in the matter dominated
era, there is a minimum wavenumber 
\begin{equation}
k_{\rm nr}\sim 0.018\,\Omega_{\rm 0m}^{1/2} \left[\frac{\Sigma
    m_{\nu}}{1\, {\rm eV}}\right]^{1/2} h/{\rm Mpc}\, ,
\end{equation}
above which the physical effect produced by neutrino free-streaming
damps small-scale neutrino density fluctuations, while modes with
$k<k_{\rm nr}$ evolve according to linear theory.  The free-streaming
leads to a suppression of power on small scales which in linear theory
can be approximated by $\Delta P / P \sim -8\,f_{\nu}$ for
$f_{\nu}<0.07$. With increasing energy content in neutrinos
(corresponding to increasing neutrino mass) the suppression becomes
larger and its shape and amplitude depends mainly on \smnu and weakly
on redshift \cite{bondefstsilk}.  At scales $k>0.1 h/$Mpc the
suppression is constant while at scales $0.01 <k (h/{\rm Mpc}) <0.1$
it gradually decreases to zero. At very large scales the effect of
neutrinos on the matter power spectrum becomes negligible. These
different scales are roughly those that are currently probed by \lya
forest data, galaxy surveys, and CMB experiments, respectively.

A large number of studies have used the effect of neutrinos on the
matter power spectrum (or perhaps better the lack thereof) to put
upper limits on the energy content and therefore the masses of
neutrinos.  Unfortunately, there is no single data set yet which fully
covers the characteristic imprint of neutrinos on the matter power
spectrum and the reliability of these limits therefore depends
strongly on the somewhat questionable assumption that there are no
systematic offsets between measurements of the matter power spectrum
with different methods which are not reflected in the quoted measurement errors.

The \lya\ forest data thereby plays a special role in probing the
effect of the free-streaming of neutrinos on the matter power spectrum
as it allows us to measure the matter power spectrum on the scales
where the suppression due to neutrinos is most pronounced while still
being in the mildly non-linear regime (\cite{croft02,vhs}, see
Ref.~\cite{meiksin09} for a more general review of the IGM).
Ref.~\cite{croftnu} have used high resolution  spectra to
obtain an early still rather weak limit of \smnu $< 5.5$ eV
from \lya forest data alone.
Ref.~\cite{seljak06}  have claimed a rather
extreme limit of \smnu $< 0.17$ eV (2$\sigma$ C.L.)
based on  the Sloan Digital Sky Survey (SDSS) quasar  data set, 
combined with other large scale structure probes. This is the
tightest limit obtained so far from cosmological data.  
Other measurements using cosmic microwave background data, galaxy redshift
surveys and growth of clusters of galaxies are usually a factor three
to six larger than this
(e.g. \cite{elgaroy02,tegmark06,mantz,komatsu09,komatsu10,reid10}).  
Note, however, also the rather low upper limit of \smnu $< 0.28$ eV
(2$\sigma$ C.L.)  obtained by \cite{shaun} based on Luminous Red
Galaxies (LRG) in the SDSS Data Release 7 combined with data on the
scale of Baryonic Acoustic Oscillations and the luminosity distance of
distant supernovae.  Forecasts for future CMB, weak lensing and \lya
forest data obtained by
Planck \footnote{http://sci.esa.int/science-e/www/area/index.cfm?fareaid=17},
the Baryon Oscillation Spectroscopic
Survey \footnote{http://www.sdss3.org/cosmology.php} and other surveys
are presented  {\it e.g.} in \cite{wang05,lespast,grattonlewis,valli09b}.

We would like, however, to stress again that the validity of current
limits depends strongly on the assumption that there are no systematic
offsets between estimates of the matter power spectrum obtained with
different methods which are not reflected in the quoted measurement errors. 
To make further progress it will be very important to identify
the characteristic signatures of the effect of neutrinos on the detailed shape 
of the matter power spectrum and its evolution with redshift. The \lya\ forest data
has here again particular potential as it covers on its own a
reasonably wide redshift range. With the \lya\ forest Baryonic
Acoustic Oscillations (BAO) survey planned as part of the Sloan Digital
Sky Survey (SDSS-3) it should be possible to reach the
scales where the suppression due to neutrinos becomes scale dependent.

While linear theory is sufficient to quantify the impact of neutrinos
on large scales and on the cosmic microwave background, the non-linear
evolution of density fluctuations has to be taken into account on
smaller scales at lower redshift.  A range of numerical studies of the
effect of neutrinos on the distribution of (dark) matter has been performed some
while ago (e.g.\cite{bondefstsilk,klypin93,maber94}) with a renewed
interest in the last couple of years
(\cite{brandbyge08,brandbyge09a,brandbyge09b}). These numerical
studies of the non-linear evolution have been complemented by
analytical estimates based on the renormalization group time-flow
approach \cite{les09,saito09}, perturbation theory \cite{wong08,saito08} or
the halo model \cite{hannestad05,abazajian05}.

The use of \lya\ forest data for accurate measurements of the matter
power spectrum benefits tremendously from the careful
modeling of quasar absorption spectra with hydrodynamical simulations
(e.g. \cite{vh06}). No such modeling has yet been performed including
the effect of neutrinos. We will be closing this gap here and present
results of the modeling of \lya\ forest data in the non-linear regime
including the effect of neutrinos by using a modified version of the
hydrodynamical code {\small GADGET-3}.

Modeling the effect of neutrinos in the mass range of interest is
non-trivial due to their rather large thermal velocities.  We mainly 
focus here on an  implementation of the neutrinos 
as a separate set of particles. Ref.~\cite{brandbyge09a} have recently suggested to model the
neutrinos with a grid based approach as a neutrino fluid instead of
neutrino particles. In this approach the gravitational force due to
neutrinos is calculated based on the linearly evolved density
distribution of the neutrinos in Fourier space. This approach has the
advantage that it does not suffer from the significant shot noise on
small scales introduced by the particle representation of the fast
moving neutrinos yielding higher accuracy at scales and redshifts
where the effect of the non-linear evolution of the neutrinos is
still moderate especially for small neutrino masses. 

In addition to our particle based neutrino simulations we have also
experimented with such a grid based implementation of neutrinos. 
In this implementation the linear growth of the perturbation in the neutrino component is 
followed by interfacing the hydrodynamical code with the public 
available Boltzmann code {\small  CAMB}\footnote{http://camb.info/}.

Further advantages of such a grid based implementation of neutrinos,
aside from eliminating the Poisson noise, are the reduced requirements
with regard to memory (there are no neutrino positions and velocities
to be stored) and computational time. However, as we will demonstrate
in Section 3 for the scales and redshift of interest for the
\lya\ forest data, non-linear effects are important. Taking their
effect into account with the particle-based implementation actually
offers a somewhat higher accuracy despite the reduction of the shot
noise at the smallest scales offered by a grid based implementation of
the linear evolution of the neutrino density.

The main improvement of our work presented here compared to previous studies are
the use of full hydrodynamical simulations in a regime in which
baryons are expected to significantly impact on the matter power
(e.g. \cite{rudd}), the focus on small scales and high redshift and the 
estimate of statistical properties of the \lya flux distribution.

The outline of the paper is as follows.  In Section 2 we describe the
numerical methods and how we generate the initial conditions for the
different simulations. Section 3 quantifies the impact of the
neutrino component on the matter power spectrum. In this section we
also address the role of numerical parameters such as the initial
redshift, number of neutrino particles, Poisson noise and velocities
in the initial conditions.  Section 4 focuses on the impact of
neutrinos on two statistics of the flux distribution in \lya\ forest 
spectra,  the flux probability distribution
function and the flux power spectrum. Section 5 presents the
upper limit on the sum of the neutrino masses  that we have  obtained 
from the SDSS flux power spectrum alone. Section 6 summarizes  our conclusions.

We recall that the scales of interest for the \lya\ forest
low-resolution SDSS spectra are $k \in [0.1-2]$ $h/$Mpc, or
$k \in [0.002-0.02]$ s/km. High-resolution spectra as the UVES/Large Programme LUQAS
sample reach $k_{\rm max}=3 h/$Mpc \cite{kim04}. The results for
neutrinos and matter power spectra will be presented as a function of
wavenumber $k$ in units of $h/$Mpc, while those that refer to
(one-dimensional) flux power spectrum will be cast in terms of
s/km. The conversion between wavenumbers expressed in s/km and $h/$Mpc
is redshift dependent and is given by the factor $H(z)/(1+z)$ which for
the cosmology used below is 99, 111.5 and 123.6 km/s/Mpc, at redshifts
$z=2$, $3$, and $4$, respectively.

\section{The simulations}

\begin{table}[h]
\begin{center}
\begin{tabular}{c|c|c|c|c|c|c|c}
\hline
linear size (Mpc$/h$) & $\Omega_{\rm m}$ & N$_{{\rm dm-gas}}^{1/3}$ & N$_{\nu}^{1/3}$& PM$^{1/3}$ & $\Sigma m_{\nu}$ (eV) & $\Omega _{0\nu} (\%) $ & $z_{IC}$ \\
\hline
60 & 0.3  &512              & 512     &--     &0.15&   0.325  & 7 \\
60 & 0.3 &512              & 512     &--    &0.3&   0.65  & 7 \\
60 & 0.3 &512              & 512     &--    &0.6&   1.3  & 7 \\
60  & 0.3&512              & 512     &--    &1.2&   2.6  & 7 \\
60 & 0.3 &512              & 1024    &--    &0.6&   1.3  & 7 \\
60 & 0.3 &512              & 1024    &--     &0.15&  0.325 & 7\\
60 & 0.3 &512              &  --     &--     &--&    --  & 7 \\
60 & 0.3 &512              &  --     &--      &--&   --  & 4 \\
60 & 0.3 &512              &  --     &--      &--&   --  & 49 \\
60 & 0.3 &384              &  --     &--      &--&   --  & 7 \\
60 & 0.3 &512              & 512     &--     &0.15&   1.3  & 4 \\
60 & 0.3 &512              & 512     &--     &0.15&   1.3  & 49 \\
60 & 0.2 &512              & 512     &--    &0.6&   1.3  & 7 \\
60 & 0.4 &512              & 512     &--    &0.6&   1.3  & 7 \\
60 & 0.6 &512              & 512     &--    &0.6&   1.3  & 7 \\
512 & 0.3&512              & --      &--     &--&   --  & 7 \\
512 & 0.3&512              & --      &--     &--&   --  & 49 \\
512& 0.3 &512              & 512     &--     &0.6&   1.3  & 7 \\
512 & 0.3&512              & 512     &--      &0.6&   1.3  & 49 \\

\hline
60 & 0.3 &512              & --  & 512        &0.6&   1.3  & 49 \\
60 & 0.3 &512              & --  & 512        &0.6&   1.3  & 7 \\
60 & 0.3 &512              & --  & 512        &0.15&   0.325  & 7 \\
60 & 0.3 &512              & --  & 1024        &0.6&   1.3  & 49 \\
512& 0.3 &512              & --  & 512        &0.6&   1.3  & 49 \\

\hline

\end{tabular}
\end{center}
\caption{Summary of the most important parameters of the  hydrodynamical simulations. The simulation 
 with box size   60 Mpc$/h$ and 512 
  Mpc$/h$ (comoving) have
  been stopped at $z=1.8$ and $z=0$, respectively. The gravitational
  softening is 4 $h^{-1}$ comoving kpc for all the different matter
  species for the small boxes and 30 $h^{-1}$ comoving kpc for the
  large boxes. The particle-mesh grid is chosen to be equal to
  N$_{\nu}^{1/3}$ for which the parameter PM (Particle Mesh, see text)
  is also reported. The bottom part of the table describes the
  grid based simulations.  Several other simulations, not reported in this
  table, have been used to test the dependence on the particle-mesh
  grid, thermal neutrino velocities in the initial conditions,
  different r.m.s. values for the power spectrum amplitude, total matter content,
  time-stepping, box-size issues, number of neutrino particles in the
  initial conditions and the method proposed (the particle based and
  grid based methods are extensively discussed in the text).}
\label{table1}
\end{table}

In order to facilitate a straightforward comparison with the findings
of \cite{brandbyge08}, we have used the following cosmological model
based on cold dark matter and a cosmological constant ($\Lambda$CDM):
$n_{\rm s}=1$, $\Omega_{\rm 0m}=0.3$, $\Omega_{\rm 0b}=0.05$, $\Omega_{\rm
  cdm}+\Omega_{0\nu}=0.25, \Omega_{0\Lambda}=0.7$ and $h=0.7$
($H_0=100\,h$ km/s).  For all our simulations, we use the
hydrodynamical TreePM-SPH (Tree Particle Mesh-Smoothed Particle
Hydrodynamics) code {\small GADGET-3}, which is an improved and
extended version of the code described in Ref.~\cite{springel05}. We
have modified the code in order to simulate the evolution of the
neutrino density distribution. The neutrinos are treated as a separate
collisionless fluid, just like the dark matter. In order to save
computational time, most of our simulations assume however that the
clustering of neutrinos on small scales is negligible and the
short-range gravitational tree force in {\small GADGET}'s TreePM
scheme is not computed for the neutrino particles. This means that the
spatial resolution for the neutrino component is only of order the
grid resolution used for the PM force calculation, while it is about an
order of magnitude better for the dark matter, star and gas particles
calculated with the Tree algorithm. We also implemented memory savings
such that the number of neutrino particles can be made (significantly)
larger than the number of dark matter particles, which helps to reduce
the Poisson noise present in the sampling of the (hot) neutrino fluid.

In the grid based implementation the power spectra of the neutrino
density component is interpolated in a table produced via {\small
  CAMB} of one hundred redshifts in total spanning logarithmically the
range $z=0-49$. The gravitational potential is calculated at the mesh
points and the neutrino contribution is added when forces are
calculated by differentiating this potential.  We have checked that we
have reached convergence with this number of power spectrum estimates
and also explicitly checked that increasing the linear size of the PM
grid by a factor two has an impact below the 1\% level on the total
matter power for the wavenumbers $k<10 h/$Mpc. For the grid
simulations the starting redshift has been chosen as $z=49$, well in
the linear regime.

The initial conditions were generated based on linear matter power
spectra separately computed for each component (dark matter, gas and
neutrinos) with  {\small CAMB}
\cite{cambcode}. The total matter power spectrum was normalized such
that its amplitude (expressed in terms of $\sigma_8$) matched the
prediction by {\small CAMB} at the same redshift. After some testing
the starting redshift for most of our runs was chosen as a rather low
$z=7$ to reduce the shot noise due to the neutrino particles.  When
generating the initial conditions, we picked random phases for the
modes in $k$-space but eliminated the Rayleigh sampling of the mode
amplitudes in order to more accurately match the mean power expected
in each mode, especially on large scales. This (artificially) reduces
cosmic variance on the scale of the box, but since we are 
mainly interested in comparing power spectra at two different redshifts
(i.e.~in the relative growth), we do expect this effect to have a
negligible impact on our main results. Initial neutrino velocities are
drawn randomly  from a Fermi-Dirac distribution  (Eq.~[\ref{eqvel}]). 
We have also tested a momentum pairing scheme in the initial conditions, 
as originally suggested in \cite{maber94,klypin93}, by splitting each neutrino
particle into two particles, giving them half the original mass and
equal but opposite thermal velocities. However, we found this to have
no influence on our results.

We have used the Zel\'dovich approximation \cite{zel} to generate
initial conditions. We acknowledge that the use of a second-order
Lagrangian perturbation theory scheme as proposed by \cite{2lpt} and
used in \cite{brandbyge08} should improve the accuracy for simulations
with low starting redshifts.  As we are mainly interested here in the
relative effect due to the free-streaming of neutrinos and not an
absolute measurement of the overall amplitude of the matter power
spectrum at a given redshift this should, however, not be a concern.

We employ a simplified criterion for star formation to avoid spending
most of our computational time on the small-scale dynamics of compact
galaxies that form in our simulations. All gas particles whose
overdensity with respect to the mean is above 1000 and whose
temperature is less than $10^5\,{\rm K}$ are turned into star
particles immediately. We have shown previously that such a star
formation recipe has very little impact on \lya flux statistics
\cite{vhs,pdflya}, but speeds up the simulations considerably.  As in
our previous simulations \cite{vhs} the heating rates have been
multiplied by a factor $\sim 3$ to achieve temperatures of the IGM at
mean density at $z=2-4$ which are in better agreement with
observational data.  For the \smnu=0.6 eV case, the mass per
simulation particle at our default resolution is $2.2\times 10^7$,
$10^8$ and $5.8\times 10^6 M_{\odot}/h$ for gas, dark matter and
neutrinos, respectively.

In Table~1 we summarize the most important parameters of the main
hydrodynamical simulations that we use in this study. We stress that
the scales and redshifts probed by most of these simulations are very
different from those explored in
Refs.~\cite{brandbyge08,brandbyge09a,brandbyge09b} but we have also
run a few simulation with the same large box size to facilitate a
comparison. Most of the simulations run for this work are moderately
time consuming. For example, the f$_{\nu}$=0.13 neutrino simulation
took about 12 hours on 200 CPUs to reach $z=2$, while increasing the
number of neutrino particles by a factor eight for the same setup
required 10 hrs on 512 CPUs, meaning that it has become about two
times slower in terms of total computational expense. For comparison,
the $N_{\rm gas}=N_{\rm dm}=512^3$ simulation of the f$_{\nu}$=0.13
model with the same amplitude of the matter power spectrum took 12 hrs
on 160 CPUs, so including neutrino particles slows down the code only
by $\sim 20\%$ (all the above numbers refer to runs performed on the
HPCS system DARWIN at Cambridge University). However, the memory
requirements for storing a large number of neutrino particles are
quite demanding, and are in fact the limiting factor for simulations
with the particle based implementation of the neutrino density. Note
that the grid based simulation of the small box size simulations has
taken about 1.6 times less CPU time to run than the corresponding
particle based simulation.  The total CPU  consumption for simulations 
with the smallest neutrino mass \smnu=0.15 eV is thereby about 10\% 
larger than that for  the largest mass we investigated  \smnu=0.6 eV.

\begin{figure}[ht]
\center
\includegraphics[width=0.75\linewidth]{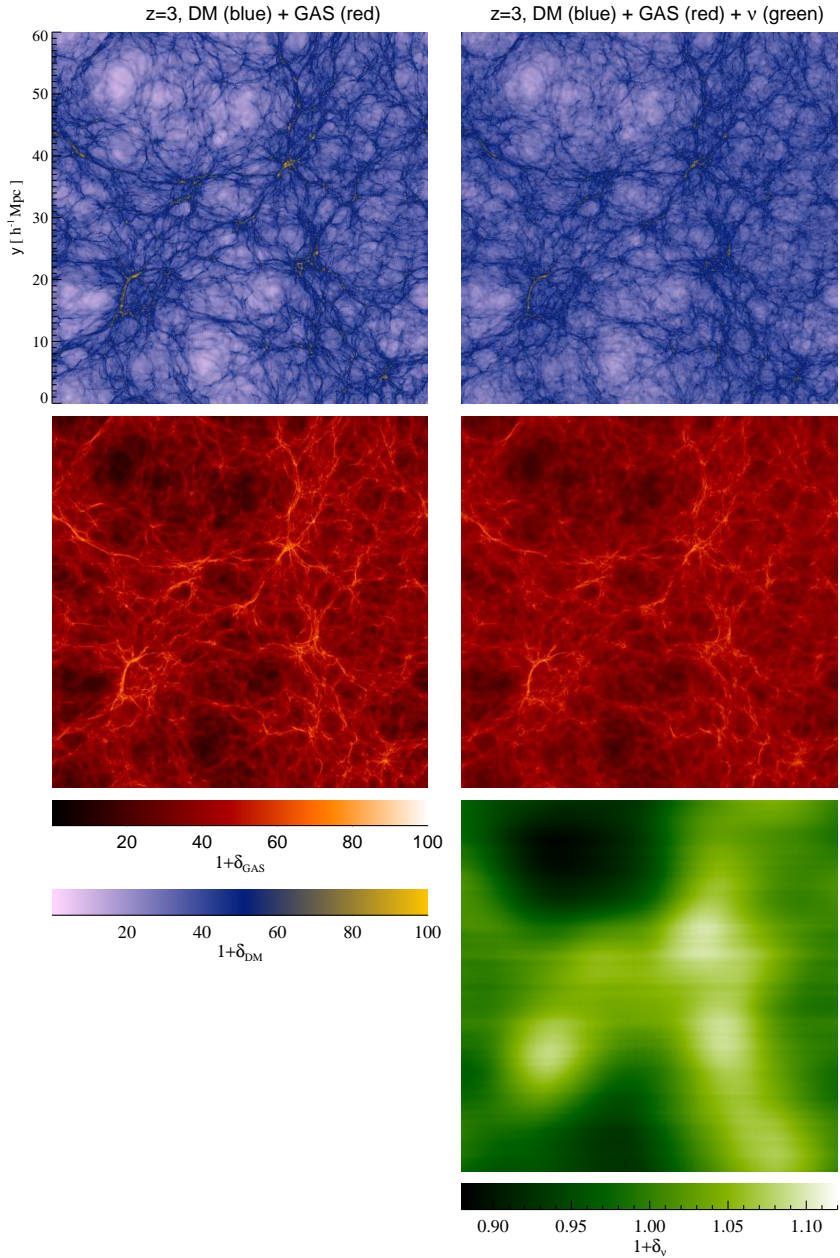}
\caption{Density slices of thickness 6 $h^{-1}$ comoving Mpc at $z=3$
  extracted from two 60$h^{-1}$ Mpc hydrodynamical simulations 
  with  gas and dark matter and no neutrinos.  The right column shows a simulation
  that includes neutrinos with \smnu=1.2 eV. The  presence of neutrinos  
  (bottom panel, green) clearly affects both the gas (red) and the
  dark  matter (blue) distribution.}
   \label{fig0}
\end{figure}

In Figure \ref{fig0} we show illustrative slices of the density
distribution of thickness 6$/h$ comoving Mpc extracted from two 60$/h$
comoving Mpc simulations at $z=3$ with and without neutrinos (for the
particle based method).  The left column shows a simulation with dark
matter and gas but without neutrinos, while the right column shows the
corresponding slices for the dark matter, gas and neutrinos for a
three-component simulation with the same initial phases and \smnu=1.2
eV.  The distribution of the neutrino density (in green, bottom panel)
has been smoothed to eliminate spurious Poisson noise at the smallest
scales in order to highlight that the genuine cosmological density
fluctuations of the neutrinos occur only on large scales due to the
free-streaming of the neutrinos. The growth of structure is clearly
less evolved in the simulation with neutrinos (the voids are for
example less empty), since the suppressed clustering of the neutrinos
slows down the growth of the perturbations in the overall matter
density. Typical neutrino fluctuations at the largest scales are about
10\% around the mean, while fluctuations of the gas and dark matter
density are usually much larger than this.

\section{The matter power spectrum}
\label{results}
\subsection{Particle based vs. grid based implementation of neutrinos} 

In a series of papers
Refs.~\cite{brandbyge08,brandbyge09a,brandbyge09b} have recently
discussed the relative benefits and drawbacks of implementing the
effect of neutrinos in the form of particles taking into account the
non-linear evolution of the gravitationally coupled neutrino, dark
matter and gas components of the matter density and a grid based
implementation accounting only for the effect of the linearly evolved
neutrino density distribution. Here, we will primarily focus on
modeling \lya\ forest data and are therefore interested in different
scales and redshifts than those probed by other authors.  However, in
order to compare our work with that of Ref.~\cite{brandbyge09a} we
performed some simulations with our grid and particle based
implementation of neutrinos with a large box size of $512
h/$Mpc. These should correspond to simulations C1 and C3 of
Ref.~\cite{brandbyge09a}.

We measure the total matter power spectrum from the simulations by
performing a CIC (Cloud-In-Cell) assignment to a grid of the same size
as the PM grid used to compute the long-range gravitational
forces. The smoothing effect of the CIC kernel is deconvolved when the
density field at the grid points is obtained. Power spectra are
computed for each component separately (gas, dark matter, stars and
neutrinos), as well as for the total matter distribution.

\begin{figure}[ht]
\center
\includegraphics[width=1\linewidth]{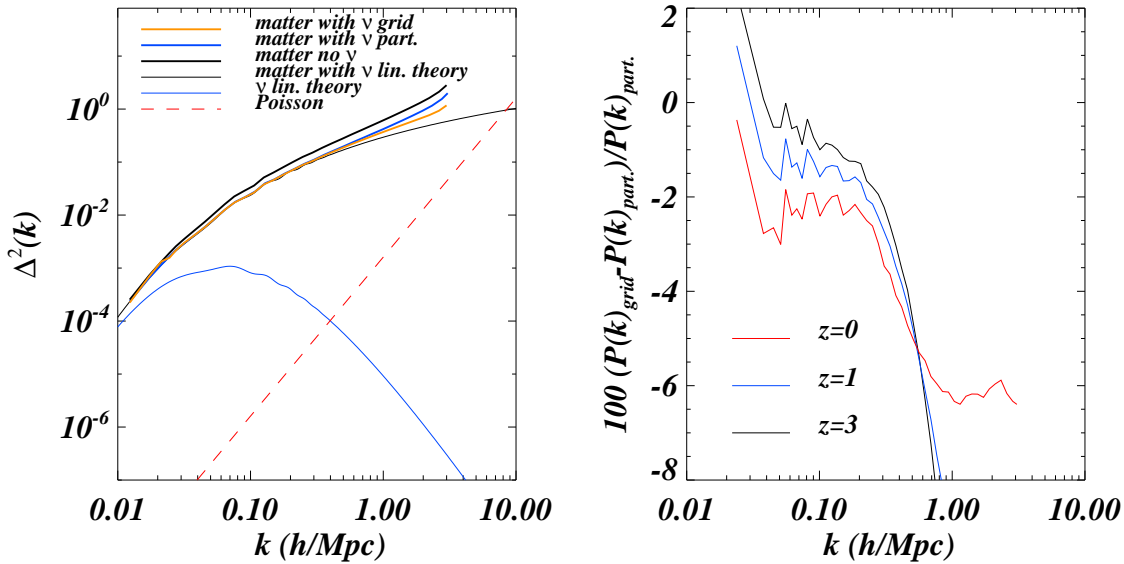}
\caption{{\it Left:} Dimensionless matter power spectrum   at
  $z=3$. We show the following quantities: linear matter power
  spectrum for a model with massive neutrinos with \smnu=0.6 eV (thin
  black line); non-linear matter power spectrum obtained with the
  particle implementation (thick blue curve) and with the grid
  implementation (thick orange curve); non-linear matter power spectrum
  for a model without neutrinos (thick black line); linear neutrino
  power spectrum (thin blue curve); Poisson contribution
  due to neutrinos (dashed red curve). All  results are for simulations with  
  box size $512$ Mpc$/h$. $N_{\nu}=512^3$
  for the particle based and $PM=512^3$ for the grid based implementation of neutrinos. {\it
  Right:}  Fractional difference of the matter power spectrum for simulations with the grid 
  and particle based implementation of neutrinos  at different redshifts ($z=0,1,3$ shown as the red,
  blue and black curves, respectively) for the large box size
  simulations with a starting redshift $z=49$.}
  \label{figcompare1}
\end{figure}

\FIGURE
{\includegraphics[width=10cm]{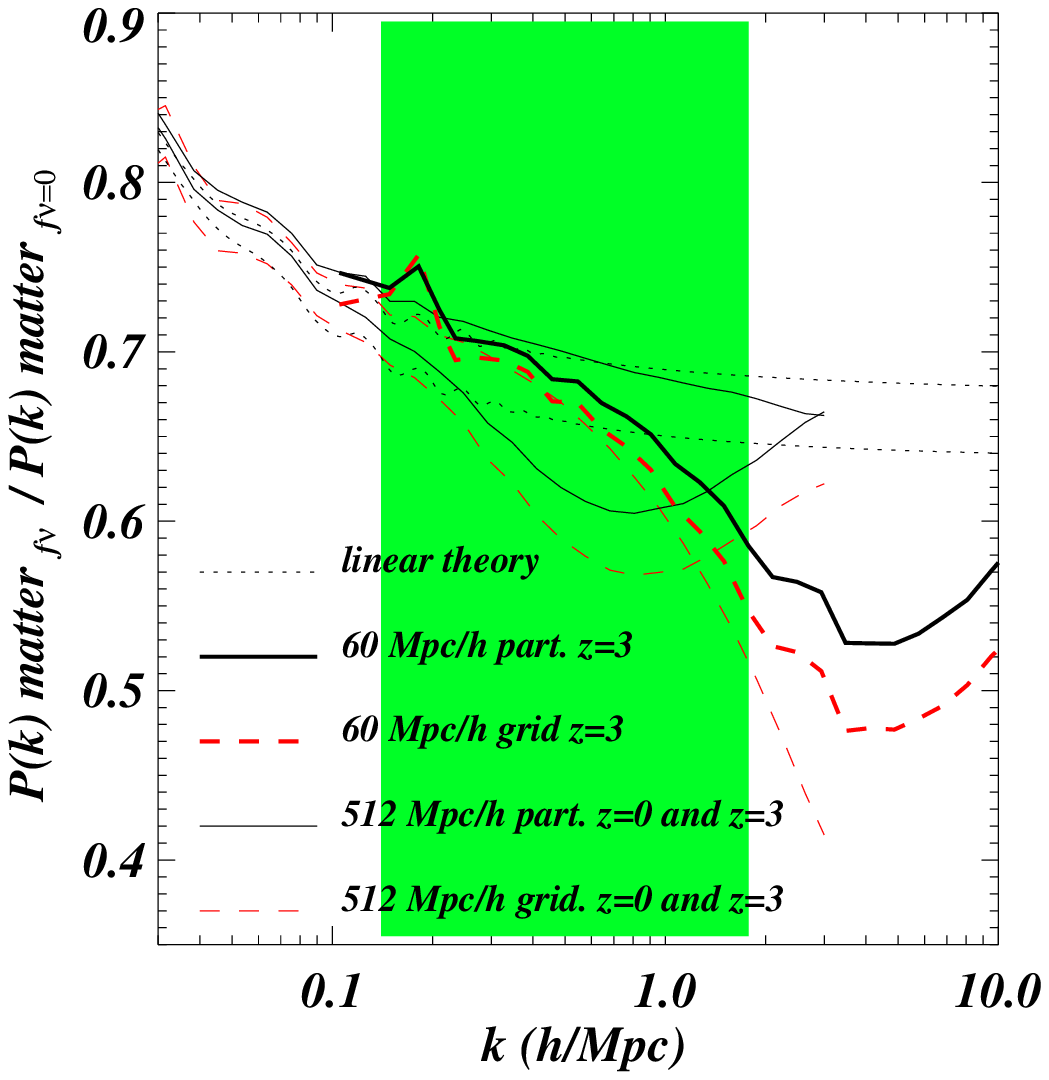}
\caption{{\it Comparison between the particle based and grid based
    implementation of neutrinos for simulations with large and small
    box size}. Ratio of matter power spectra for simulations with and
  without neutrinos as described in the text.  The thin curves refer
  to simulations with a large linear box size ($512/h\, $Mpc): the
  grid based neutrino implementation (thin red dashed curves) and the
  particle based neutrino implementation (thin black continuous
  curves) at $z=0$ and $z=3$. The thick curves refer to simulations
  with the default linear box size of $60/h\,$Mpc: with the grid based
  (thick red dashed curve) and the particle based implementation
  of neutrinos, (black continuous curve). The dotted curves show the
  predictions of linear theory at $z=0$ and $z=3$. The shaded area
  indicates approximately the scales that are probed by the SDSS flux
  power spectrum data set.
\label{figcompare2}}}

In the left panel of Figure \ref{figcompare1} we compare the matter
power spectra at $z=3$ in dimensionless units for simulations with
neutrino mass \smnu=0.6 eV for the grid based (thick orange curve) and
particle based (thick blue line) implementations with that of a
simulation without neutrinos (thick black curve) and the prediction of
linear theory (thin black curve). We also show the neutrino power
spectrum (thin blue curves) and the (redshift independent) Poisson
contribution (red dashed curve). The Poisson contribution (for the $N_{\nu}=512^3$ case) exceeds the
neutrino power at $k>0.8 h/$Mpc. Only results for simulations with the
large box size are shown. In the right panel we show the fractional
difference of the matter power spectrum of simulations with the
particle and grid based implementation of neutrinos at different
redshifts (in percent). At large scales $k<0.5 h/$Mpc the differences
are largest at $z=0$, of the order of 2 \% while at $z=3$ are smaller
and around 1\%.  The results can be directly compared to those
obtained by \cite{brandbyge09a} for the same \smnu= 0.6 eV (figure 1
in their paper) but note that despite our attempt to choose similar
parameters there may be still small differences in some of the
parameters and that the simulations in \cite{brandbyge09a} do not
contain baryons. The discrepancies between the two implementations
albeit small on large scales appear to be somewhat larger in our
simulations.

In Figure \ref{figcompare2} we compare results from the two methods in
terms of neutrino suppression for the large simulation box with
results for a box size nearly ten times smaller ($60/h$ Mpc), more
appropriate for the modeling of \lya forest data. Large boxes are
shown as thin curves which are red dashed in the grid implementation
and black continuous in the particle one, respectively. Smaller boxes
are reported as thick curves only at $z=3$.  At the smaller scales,
that are not fully resolved by the large box-size simulation,
non-linear effects are already important at the redshifts probed by
\lya forest data and this is clearly demonstrated by the discrepancies
between large and small scales. In fact, at $(k=k_{\rm max},z=3)$
$\Delta^2_{\rm non linear}\sim 3 \Delta^2_{\rm linear}$ (see left
panel of Figure \ref{figcompare1}) and this non-linear evolution is
missed in the large box simulations.  We have checked that we get
numerical convergence in terms of non-linear matter power spectra
between the $N_{\rm dm,gas}=512^3$ and the $N_{\rm dm,gas}=384^3$
cases, so our results can be trusted at a quantitative level. We
interpret this discrepancy as due to the fact that in simulations with
the grid-based implementation the enforced linear evolution of the
neutrinos with the same phases prevents a proper response to the dark
matter growth.  At the small scales Fourier mode mixing is important
for the phase association and can alter the linear theory picture
significantly.  This appears to result in a significantly larger
discrepancy between simulations with the grid and particle based
implementations on the scales and redshifts relevant for \lya forest
data.  The  differences between should thereby be  mainly due to the
fact that the non-linear evolution at small scales is not properly
reproduced by the grid method.
We will therefore focus mainly on simulations with the particle
based implementation in the rest of the paper, keeping in mind that
our results are affected by Poisson noise in the neutrino components
at the smallest scales probed.

Note that increasing the accuracy of the simulations with the particle
based neutrino implementation further by pushing up the number of
neutrino particles in order to decrease the Poisson contribution to
the matter power spectrum is rather demanding in terms of parallel
computing resources.  Increasing our default number of neutrino
particles ($512^3$) by a factor of eight is still doable on the
machine we had available for this (DARWIN) and requires a factor $\sim
2$ more CPU time.  Increasing the default number of neutrino particles
by a factor of 27 instead resulted, however, in prohibitive memory
requirements.  Further improvements in the accuracy of the simulations
for future more accurate data sets will thus probably require
optimization of the necessary compromises in a hybrid of the grid and
particle based implementations as proposed by \cite{brandbyge09b}.

\subsection{The effect of the neutrinos on the matter power spectrum}

In this Section we first contrast the effect of the free-streaming of
neutrinos on the power spectrum of the total matter density for a
range of neutrino masses in full numerical simulation with that
predicted by linear theory. We will only refer to results from simulations 
with the particle based implementations of neutrinos  here, unless explicitly stated. 
In order to quantify the
suppression of structure growth induced by neutrinos, we divide the
matter power spectra of the neutrino simulations by the corresponding
matter power spectrum extracted from the $\Lambda$CDM simulation
without neutrinos.

\FIGURE
{\includegraphics[width=15.5cm]{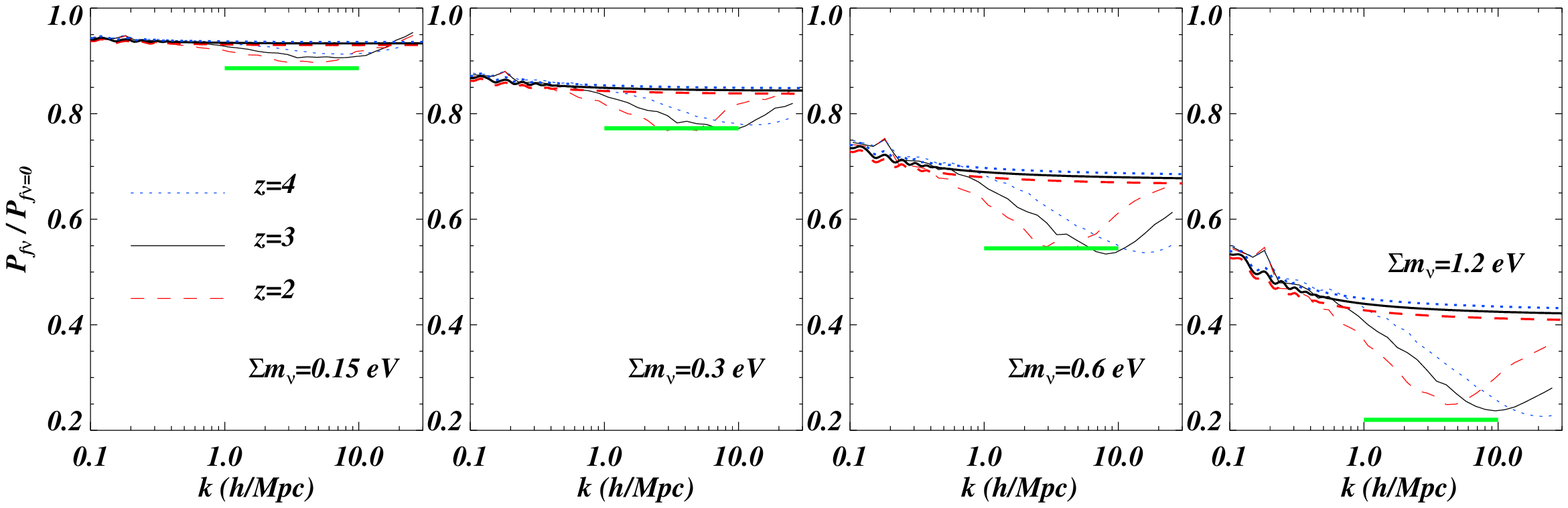}
\caption{{\it Effect of different $f_{\nu}$ on the matter power and
    comparison with linear prediction}. Ratio between matter power
  spectra for simulations with and without  neutrinos
  for four different values of the neutrino mass, 
  $\Sigma m_\nu =0.15$, 0.3, 0.6, $1.2$~eV, from left
  to right. Different line-styles refer to different redshifts: $z=2$ (red
  dashed), $z=3$ (black continuous) and $z=4$ (blue dotted). 
  The predictions of linear theory are shown as the thick
  curves. An estimate of the overall suppression based on 
  the hydrodynamical simulations is shown as a thick short green line,
  $\Delta P/P \sim -10.5\, f_{\nu}$.}
   \label{fig2}}

In Figure~\ref{fig2}, we compare the non-linear power spectra from the
numerical simulations with the results predicted by linear theory,
shown as thick curves.  The suppression of the matter power spectrum
increases with increasing $\Sigma m_{\nu}$ (recall that these
simulations are normalized at the CMB scale).  Note the plateau of
constant suppression predicted by linear theory, which is
approximately described by $\Delta P/P \sim -8\, f_{\nu}$, and depends
only very weakly on redshift.  Linear theory provides a good
description of the matter power spectrum at $z=2-4$ up to scales of
about $k\sim 0.4$ $h/$Mpc, and the agreement is more accurate for the
smaller neutrino masses. The non-linear matter power spectrum does, on
the other hand, depend strongly on redshift and the dependence on
scale becomes steeper with decreasing redshift. For \smnu=0.6~eV, a
good fit to the suppression at $z=3$ in the range that deviates from
linear theory, $k \,(h/{\rm Mpc}) \in [0.3,3]$, is given by
$P_{f_{\nu}}/P_{f_{\nu}=0}=T_{\nu}(k) \propto {\rm
  log}_{10}(k)^{-0.15,-0.11,-0.08}$ at $z=2,3,4$, respectively; while
for \smnu=0.3~eV, we find $T_{\nu}(k) \propto {\rm
  log}_{10}(k)^{-0.08,-0.06,-0.04}$ at the same redshifts. We also
note that the maximum reduction of power shifts to larger scales with
decreasing redshift.

The maximum of the  non-linear suppression can be 
described by $\Delta P/P \sim -10.5 \, f_{\nu}$ (green thick curves in
Fig. \ref{fig2}) for neutrino masses \smnu=0.15, 0.3, 0.6~eV,
respectively. For the most
massive case we considered the suppression is about $\Delta P/P \sim -9\,
f_{\nu}$. Our results differ somewhat  from those of 
Ref.~\cite{brandbyge08}, who reported $\Delta P/P \sim -9.8 \, f_{\nu}$ (at
$z=0$) while we measure $\Delta P/P \sim -9.5 \, f_{\nu}$, apart from the most
massive case in which the suppression is smaller,   $\Delta P/P
\sim -8\, f_{\nu}$. We must remind,  however,  that the above linear approximation starts to
break down for large neutrino masses and is already very poor for \smnu=1~eV
(e.g.~\cite{lespast}).

\FIGURE
{\includegraphics[width=15.5cm]{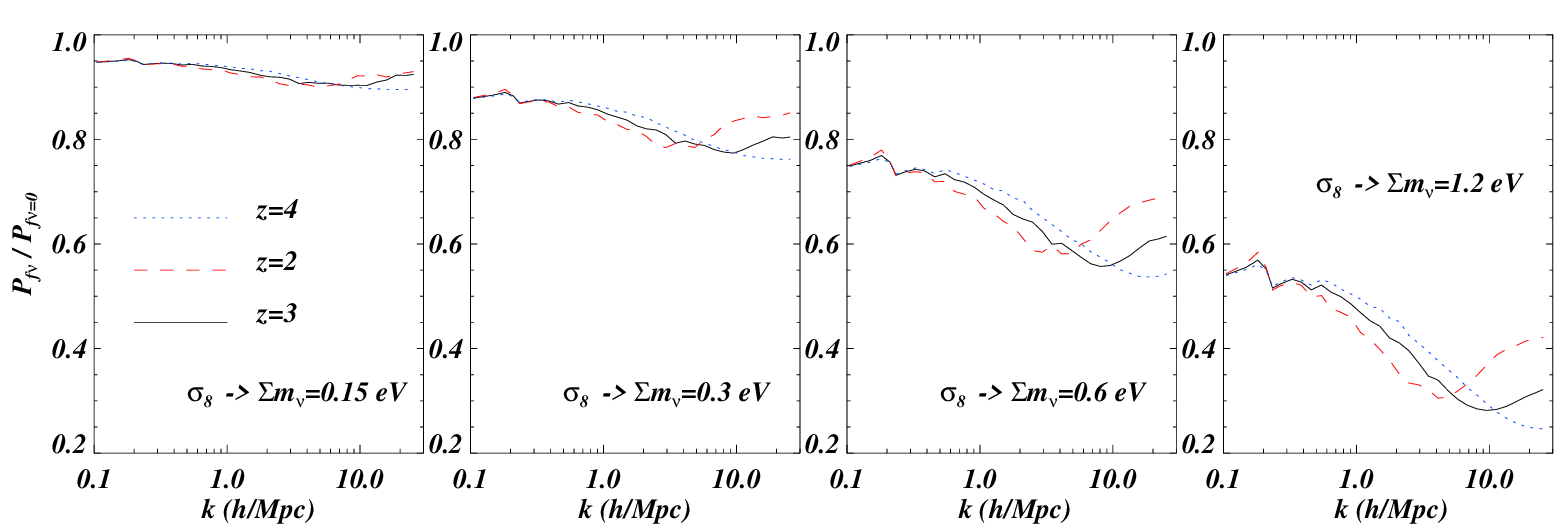}
\caption{{\it Effect of a different r.m.s. value for the amplitude of the
    matter power spectrum.} Ratio between matter power spectra
  with different values of $\sigma_8$. Four different cases are presented that
  have exactly the same $\sigma_8$ at $z=7$  as those of the models with $\Sigma m_\nu
  =0.15$, 0.3, 0.6, $1.2$~eV, from left to right. 
   The different line-styles  refer to
  different redshifts: $z=2$ (red dashed), $z=3$ (black continuous),
  and $z=4$ (blue dotted).}
   \label{fig3}}

\FIGURE
{\includegraphics[width=15.5cm]{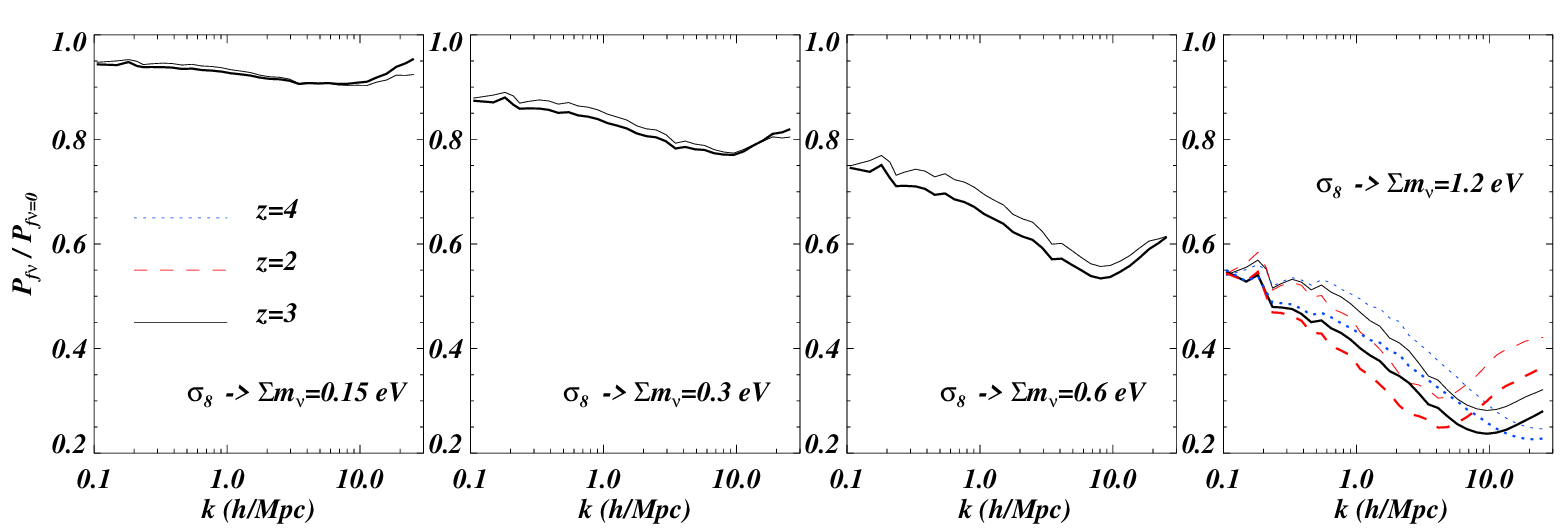}
\caption{{\it Effect of $\sigma_8$ vs. effect of $f_{\nu}$}. Ratio between
  matter power spectra with different values of $\sigma_8$ (no neutrinos, thin
  curves) and different neutrino energy density (with neutrinos, thick curves). Four
  different cases are presented that have exactly the same $\sigma_8$
  at $z=7$  as the models including neutrinos with 
  $\Sigma m_\nu =0.15$, 0.3, 0.6, $1.2$~eV from left to
  right. Different line-styles  refer to different redshifts: $z=2$ (dashed),
  $z=3$ (continuous), and $z=4$ (dotted). For clarity we show the
  three different redshifts only for the most massive case (rightmost panel).
  In the other panels we show the $z=3$ results only.}
   \label{fig4}}

Note that there is an up-turn in the suppression at scales of about $5$, $8$, $10$
~$h$/Mpc for $z=2$, $3$, $4$, respectively. A similar  upturn was  found by
Ref.~\cite{brandbyge08}, but at a scale of $1~h/$Mpc at $z=0$.  We
have checked that this feature  does not
depend on the number of neutrino particles in the simulation. It does depend weakly on
the value of $f_{\nu}$ (or $\Omega_{\rm cdm}$), moving to larger scales when
$f_{\nu}$ is decreased. The upturn  appears to be related to the non-linear
collapse of haloes, which decouple from the large scale modes slightly
differently in simulations with neutrinos than in simulations that have a
different value for the amplitude of the power spectrum and no neutrinos. This
suggests that the virialization of halos is slightly modified by the smoothly
distributed neutrino component, in a similar fashion as done by dark energy
where this is a well-known effect (see \cite{alimi} for a recent study).

The main effect of the free-streaming of neutrinos is a reduction of
the amplitude of the matter power spectrum on small scales.  This
results in a well known degeneracy between the values of $\sigma_8$ and \smnu.  
In order to explore this degeneracy in more detail we have
run four further hydrodynamical simulations without neutrinos that
have the same value of $\sigma_8$ at $z=7$ as the four different
neutrino simulations, namely: $\sigma_8$ = 0.137, 0.132, 0.122, and
0.103, mimicking the simulations with  \smnu$=0.15$, $0.3$, $0.6$, and $1.2$~eV 
(note that the simulation without neutrinos has $\sigma_8=0.141$ at
$z=7$). The corresponding $z=0$ values are $\sigma_8 = 0.845$,
$0.806$, $0.732$, $0.611$, and $\sigma_8=0.878$ for the default
simulation.  The differences in terms of the amplitude of density
fluctuations range from 3\% (0.15 eV) and 36\% (1.2~eV). For the
0.6~eV simulations  the difference is 15\% which is very close to the
corresponding $1\sigma$ uncertainty in the linear matter power
spectrum amplitude at $z=3$ at scales $k=0.009$ s/km, as derived from
SDSS \lya observations by \cite{mcdonald05}.

The results of the simulations without neutrinos but  a decreased
power spectrum amplitude are shown in Figures~\ref{fig3} and
\ref{fig4}, where we can see  that the effects of neutrinos and a 
overall suppression of the matter power spectrum amplitude  are very similar. 
For \smnu=0.15, $0.3$~eV the  differences are at the
percent level, while for the simulations with more massive neutrinos 
they differences increase to about 5-10\%. In the 
rightmost panel of Fig.~\ref{fig4} (\smnu=1.2 eV) , where the 
suppression is largest, we show the results of the neutrino
simulations of Figure~\ref{fig2} for all three redshifts.  
On the small scales considered here the effect of the neutrinos on the  
non-linear matter power spectrum is to a high degree degenerate with an
overall reduction of the matter power spectrum amplitude. The scale
and redshift dependent differences to a simulation with an overall
decrease of the power spectrum amplitude are small but nevertheless noticeable and
increase with increasing neutrino mass. For neutrino masses with  \smnu
$> 0.6$ eV, the simulations with neutrinos typically show an
additional suppression at the 5-10\% level when compared to simulations 
without neutrinos with the same $\sigma_8$ at $z=7$. Note that Figure~\ref{fig4} is meant to 
highlight the small differences in the shape of the matter power spectra due to
neutrinos if   a normalization of the matter power spectrum at small 
scales is assumed (same $\sigma_8$, i.e.~the fluctuations are effectively normalized at the \lya
forest scale).

\subsection{The effect of  varying  the total matter content}

In the last section we had investigated the effect of varying the neutrino mass 
at fixed total matter content $\Omega_{\rm m}$. In linear theory  the effect of neutrinos 
is well parameterized by the ratio of mass content in neutrinos to total matter content, 
$f_{\nu}$. We test here how well this holds for full non-linear simulations including 
neutrinos by varying 
 $\Omega_{\rm m}$ at fixed neutrino mass.  We have run
simulations with and without neutrinos with  $\Omega_{\rm
  m}=0.2,0.4,0.6$  and  \smnu=0.6 eV.  The  results are shown 
  in Figure \ref{figomega}.  To further test the  degeneracy with 
  simulations with adapted value of $\sigma_8$  we also run two simulations
without neutrinos but with the same $\sigma_8$ value as the simulations 
with $\Omega_{\rm m}=0.2,0.4$ . We  overplot the results as thick
curves.  Note that this will  mimic  results obtained with  simulations 
without neutrinos particles  where the effect of neutrinos has been 
approximated by  changing the (initial) power spectrum of the other matter 
components. This approximate approach has been taken by most studies 
in the past.

\FIGURE
{\includegraphics[width=15.5cm]{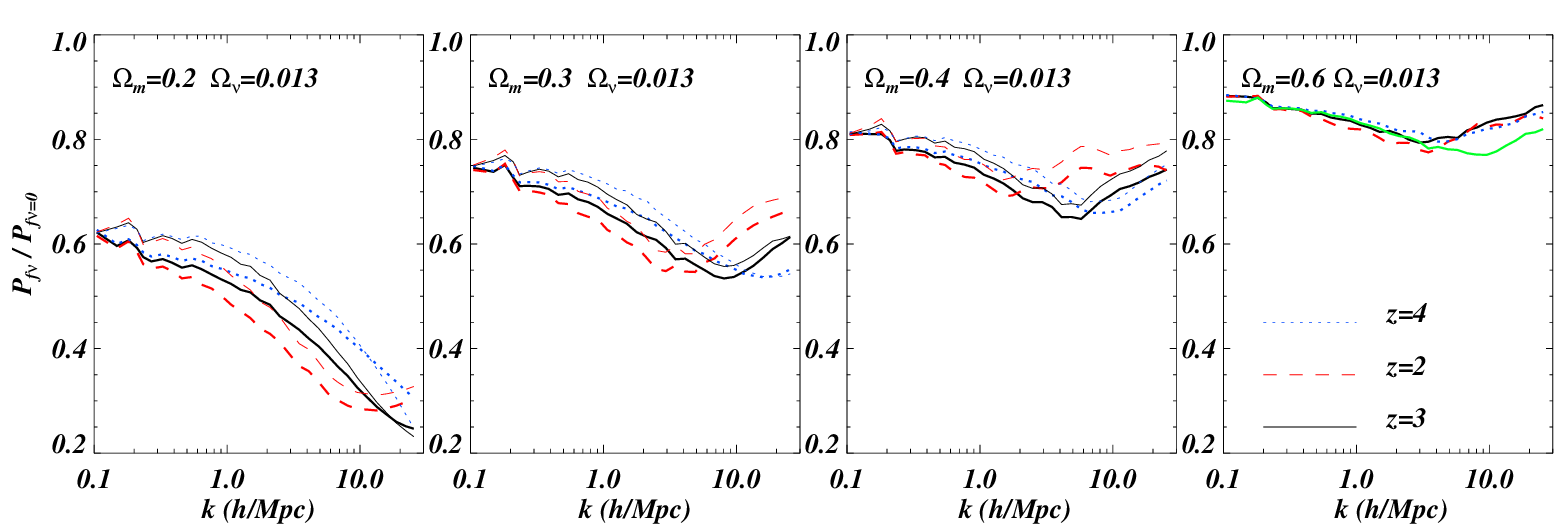}
\caption{{\it Effect of varying  $\Omega_{\rm m}$ at fixed
    \smnu}. Ratio between matter power spectra with different values
  of $\Omega_{\rm m}$ at fixed neutrino contribution to the 
  energy density (\smnu=0.6eV). Four different cases are shown  with $\Omega_{\rm
    m}=0.2,0.3,0.4,0.6$, from left to right. Different line-styles
  refer to different redshifts: $z=2$ (dashed), $z=3$ (continuous),
  and $z=4$ (dotted). The thick curves in the first three panels
  are for simulations without neutrinos but with the same
  $\sigma_8$ as the  simulations with neutrinos. The green thick curves  
  in  the  right-most panel are for simulation with   $(\Omega_{\rm m}=0.3,\Omega_{\nu}=0.0065)$ 
   that has the same  $f_\nu$ as the $(\Omega_{\rm m}=0.6,\Omega_{\nu}=0.013)$ one.  }
   \label{figomega}}

We note the same trends as before. The presence of neutrinos
results in an additional  suppression of  the matter power that,
is  well parameterized by the quantity $f_\nu=\Omega_\nu/\Omega_{\rm m}$
also in the non-linear regime. At fixed neutrino mass 
the suppression is  therefore larger for smaller values of $\Omega_{\rm m}$.
For example,  for $\Omega_{\rm m}=0.2$  it  is twice  larger than for 
$\Omega_{\rm m}=0.4$  and of about the same level of
$-10.5\,f_\nu$ that we found for the fixed $\Omega_{\rm m}=0.3$ case.  We
thus conclude that the dependence of the matter power spectrum in the
non-linear regime on the quantity $\Omega_{\rm m}$ is also well
captured by the parameter $f_\nu$. This is also demonstrated by the
green line in the right-most panel of Figure \ref{figomega} where the
non-linear power spectra for the
$(\Omega_{\rm m}=0.6,\Omega_{\nu}=0.013)$ are overplotted on  the $z=3$
result obtained for  $(\Omega_{\rm m}=0.3,\Omega_{\nu}=0.0065)$: the
agreement is almost perfect at the scales of interest here, showing
that also in the non-linear regime the $f_{\nu} $ parameter is the
relevant quantity to describe the effect of neutrinos.   We therefore conclude that 
on scales relevant for the Lyman-$\alpha$ forest data  the parameter $f_\nu$ is sufficient 
to characterize the effect  of  $\Omega_{\rm m}$ and  the neutrino mass 
also on  the non-linear  power spectrum.
although these results should be confirmed by using larger box size
simulations, where possibly one starts to be more sensitive to the
overall shape of the matter power spectrum.

\subsection{Resolution tests and dependence on the initial conditions}

In this subsection we investigate several numerical effects that
impact the power spectra measurements presented in the previous
section: the number of neutrino particles, the velocities in the
initial conditions, the sampling of the initial conditions with
neutrino pairs to balance momentum, and the starting
redshift. Here we are not discussing resolution effects with
regard to the number and mass of dark matter and gas particles since  these
have already been discussed extensively elsewhere with respect to the
scales probed by \lya data (e.g. \cite{vhs,mcdonald05,vielhpm}). 
Furthermore, our results are mainly presented in terms of ratios of power
spectra extracted from simulations with the same resolution. This
strongly reduces the sensitivity to the dark matter and gas
resolution.

\FIGURE
{\includegraphics[width=15.5cm]{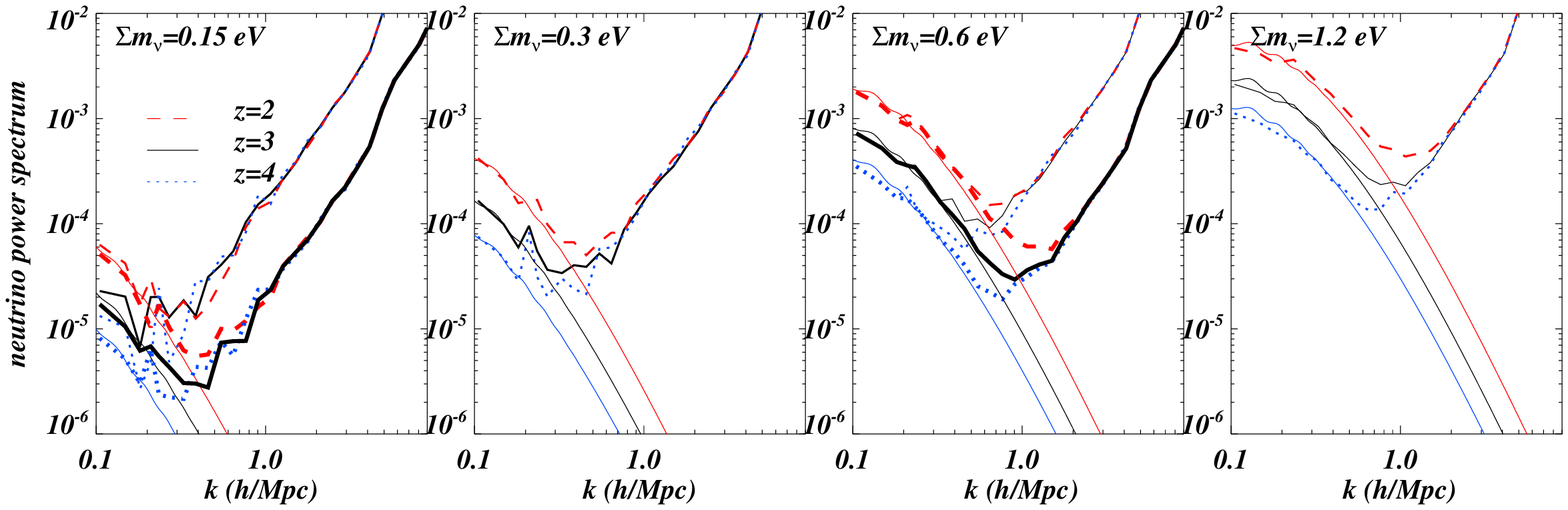}
\caption{{\it Effect of different $f_{\nu}$ on the neutrino power spectrum and
    comparison with prediction of linear theory}. Power spectra 
  (dimensionless units) for the neutrino
  component as a function of wavenumber. Four different cases are presented
  with $\Sigma m_\nu =0.15$, 0.3, 0.6, $1.2$~eV from left to right. Different
  line-styles refer to different redshifts: $z=2$ (red dashed), $z=3$ (black
  continuous), and $z=4$ (blue dotted ). The prediction of linear
  theory are represented by the continuous thin curves. In the first and
  third panel, for the $\Sigma m_\nu =0.15$, $0.6$~eV cases, respectively, we
  also show the simulations with eight times more neutrino
  particles, i.e.~$N_{\nu}=1024^3$ instead of
  $N_{\nu}=512^3$ particles (thick curves).}
   \label{fig5}}

\FIGURE
{\includegraphics[width=14cm]{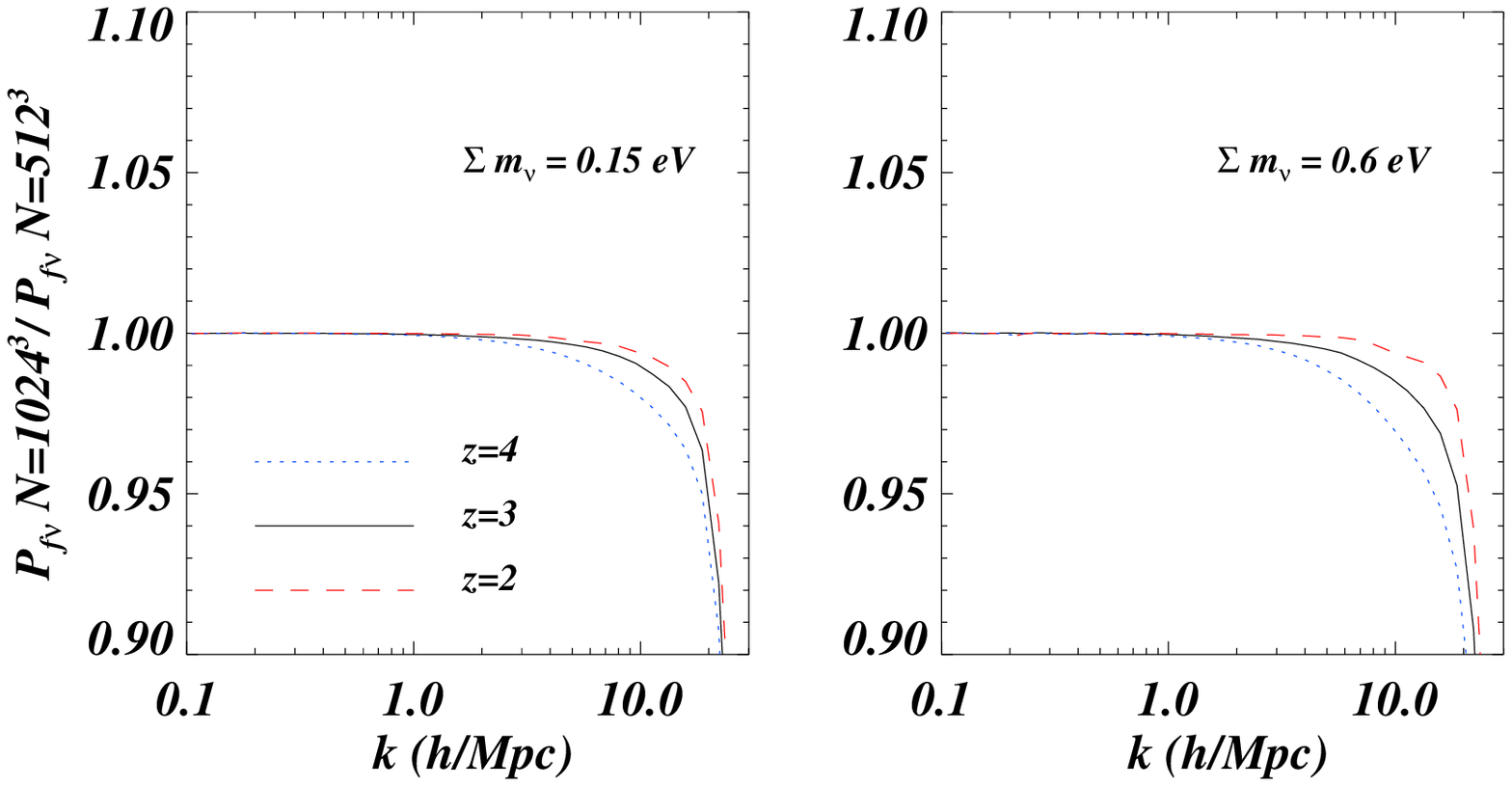}
\caption{{\it Resolution test for simulations with neutrinos: effect
    on the matter  power spectrum}. Ratio between matter power spectra with different number of
  neutrino particles ($512^3$ and $1024^3$) for simulations with  $\Sigma
  m_\nu =0.15$ eV (left panel) and $\Sigma m_\nu =0.6$ eV (right
  panel). Different line-styles refer to different redshifts: $z=2$
  (red dashed),  $z=3$ (black continuous), and $z=4$ (blue dotted).}
   \label{fig6}}

First, we  will take a closer look at the power spectrum of the
neutrino component of the matter density.  In Figure~\ref{fig5}, we
compare the non-linear neutrino power spectrum with predictions from
linear theory for some of the simulations in  Table 1. At scales 
of about $\sim 1\, h/$Mpc the power  spectrum
starts to deviate strongly from linear theory and follows instead the
expectation for Poisson noise, $P(k)\propto k^3  L_{\rm box}^3/N_{\rm
particles}$. The Poisson contribution to the power
spectrum  depends as expected on the number of neutrino particles
used. This is  demonstrated in the first and third panels where we also show
results for simulations with \smnu=0.15, $0.6$~eV and $N_{\nu}=1024^3$ 
instead of $512^3$ neutrino particles.  Doubling the number of neutrino
particles for each spatial dimension shifts the Poisson contribution to the
matter power spectrum by a factor of roughly two to smaller scales. 


When modeling the \lya forest flux power spectrum  one ideally would like to
sample the neutrino power spectrum properly on scales  between 0.1 and
2 $h/$Mpc. As evident from  Figure~\ref{fig5} this will be
difficult as the neutrino distribution is affected by shot noise  
at the smallest relevant scales. Reducing this shot noise to
negligible levels requires a number of neutrino particles with 
memory requirements beyond our current capabilities. In the
following, we will see that despite the fact that
the neutrino power spectrum is affected by shot noise at the smallest 
scales relevant for \lya studies, the impact on the one-dimensional flux
power is still very small.

\FIGURE
{\includegraphics[width=10cm]{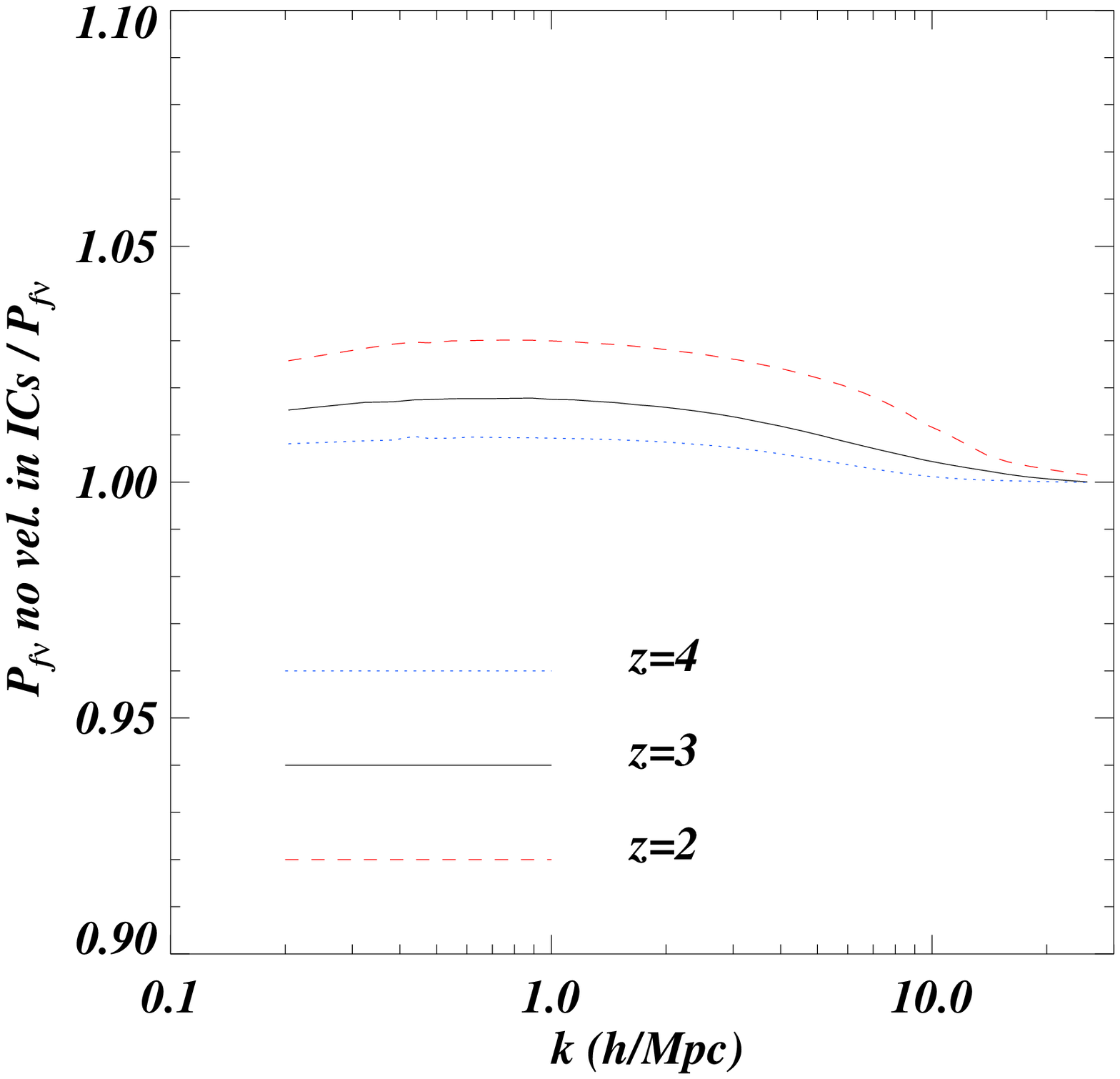}
\caption{{\it Effect of neutrino velocities on matter power spectra}. Ratio
  between matter power spectra with and without considering the velocities in
  the initial conditions at $z=7$ ($\Sigma m_\nu
  =0.6$~eV). Different line-styles refer to different redshifts: $z=2$
  (dashed red), $z=3$ (continuous black), and $z=4$ (dotted blue).}
   \label{fig7}}

In Figure~\ref{fig6}, we show the ratios of the (total) matter power spectra
for a simulation with  $N_{\nu} = 1024^3$ neutrino particles to that
of our default simulations with $512^3$ neutrino particles, 
for \smnu = 0.15~eV (left panel) and \smnu = 0.6~eV (right
panel). The increased number of neutrino particles results in  an
additional suppression of about 5-10\% at scales above $k\sim 10 h/$Mpc, while at the
scales probed by the \lya forest the effect on the total matter power is
of the order of 1\% or less. 
While the  Poisson contribution to the
neutrino power spectrum 
is significant, its effect on the the total matter power spectrum is
still small at small scales. The suppression is thereby slightly larger for  \smnu = 0.6 eV,
where the neutrinos constitute a larger fraction of the  overall matter
density.  However, we should  stress here that the \lya data is primarily
sensitive the  one-dimensional matter distribution along the
line-of-sight (although cross-correlating information in the
transverse direction is a promising tool for future observational data
sets). The one-dimensional power spectrum, being a projection of the
three-dimensional information, will be affected out to larger scales than the
three-dimensional power spectrum by the suppression (or increase) of power at
a given scale \cite{viel02,selsterile}.

\FIGURE
{\includegraphics[width=15cm]{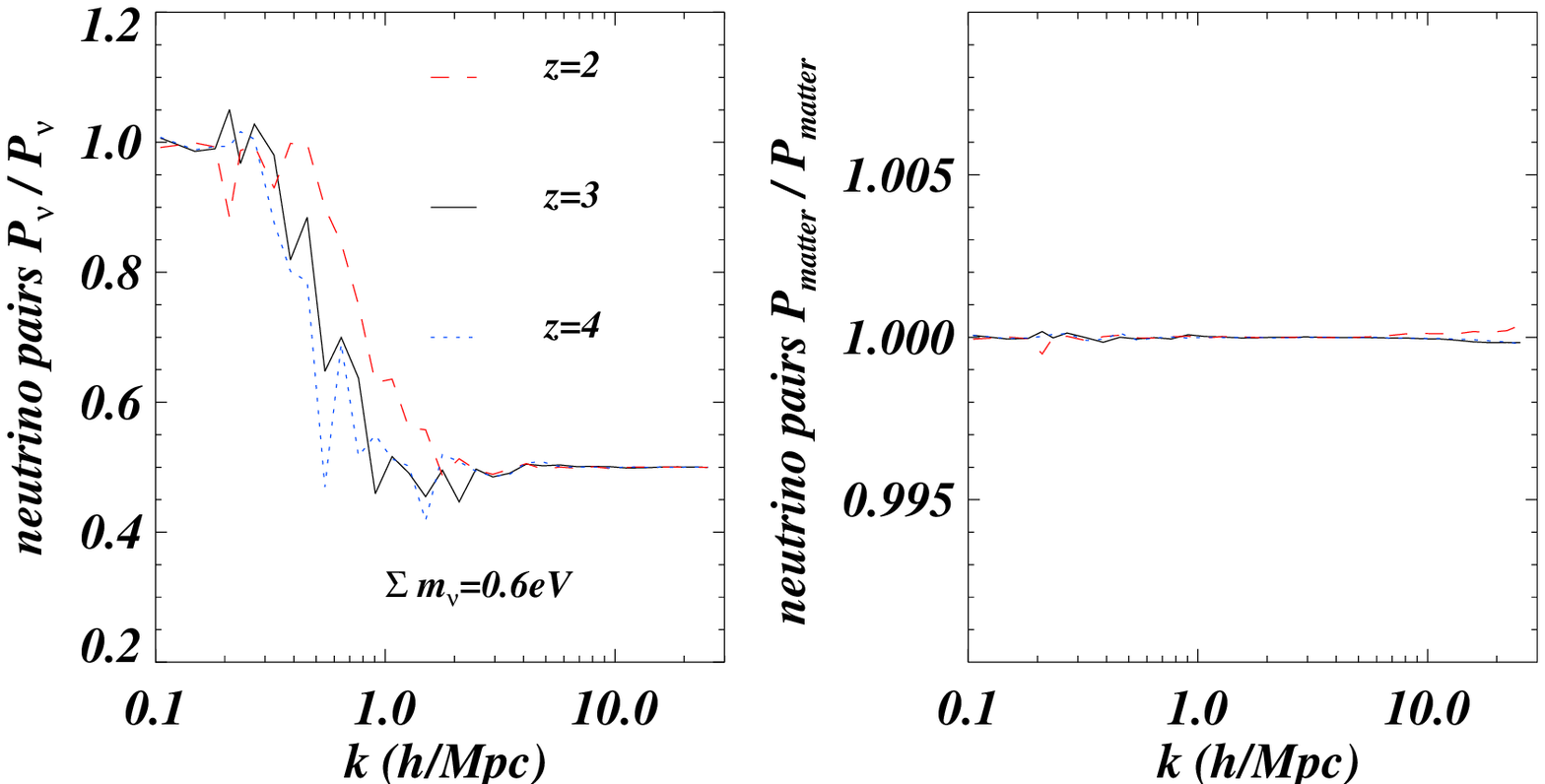}
\caption{{\it Effect of a momentum conserving sampling of the initial
    distribution using neutrino pairs}. Ratio between neutrino power spectra
  (left panel) and matter power spectra (right panel), with and without
  pairing of neutrinos in  the initial conditions at $z=7$ ($\Sigma
  m_\nu =0.6$~eV). Different line-styles refer to different
  redshifts: $z=2$ (dashed red), $z=3$ (continuous black), and $z=4$
  (dotted blue).}
   \label{fig8}}

\FIGURE
{\includegraphics[width=9cm]{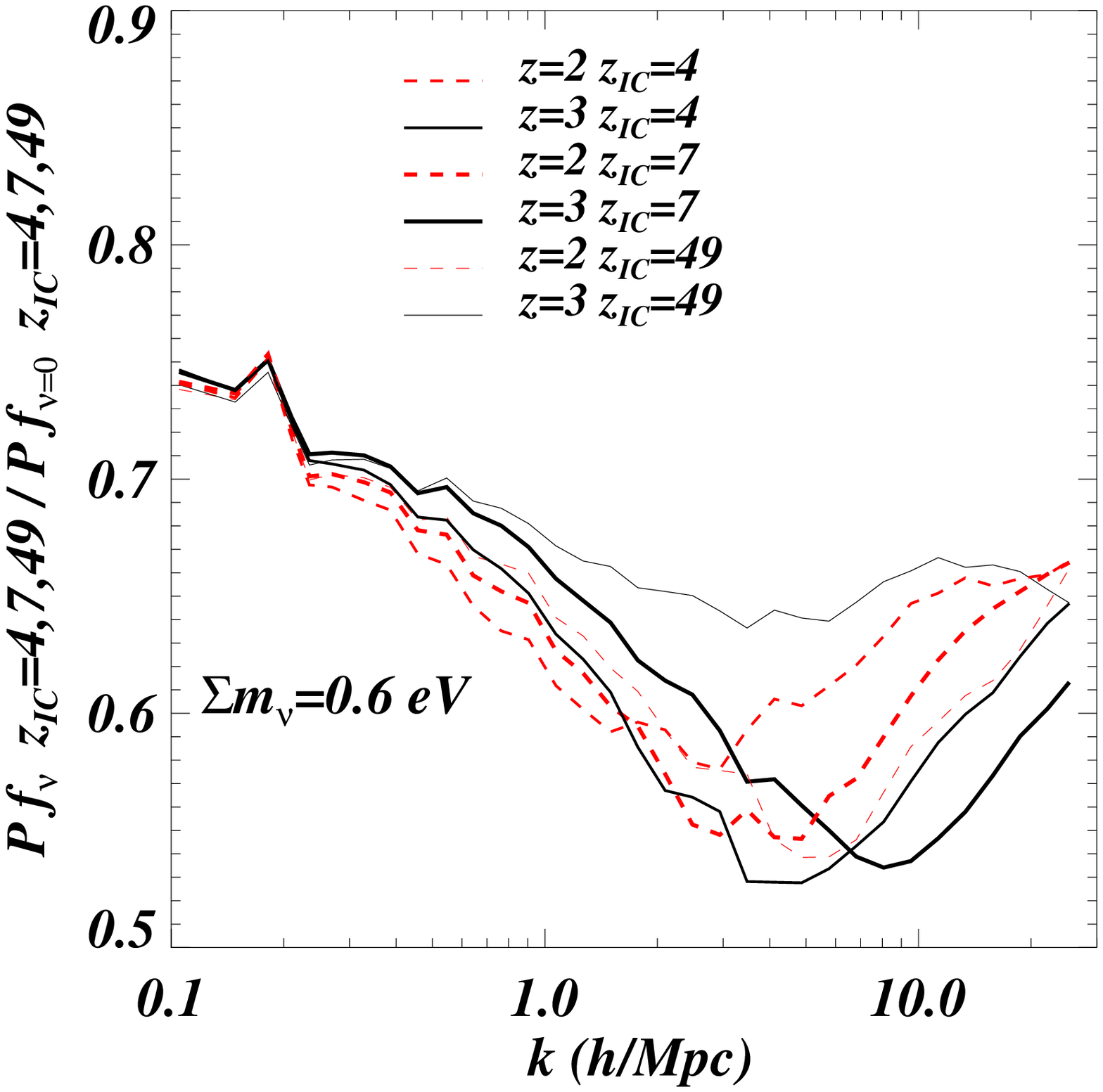}
\caption{{\it Effect of the initial redshift on the matter power spectrum}. Comparison of
  the ratio between the matter power spectra of simulations with different
  initial redshift $z_{\rm IC}$. We show the following quantities at $z=2$
  (red dashed) and $z=3$ (black continuous):
  $P(k,f_{\nu},z_{IC}=4)/P(k,f_{\nu}=0,z_{IC}=4)$ (thin curves),
  $P(k,f_{\nu},z_{IC}=7)/P(k,f_{\nu}=0,z_{IC}=7)$ (very thick curves and
  default case) $P(k,f_{\nu},z_{IC}=49)/P(k,f_{\nu}=0,z_{IC}=49)$ (thick
  curves). All simulations are for  $\Sigma m_{\nu} =0.6$ eV.}\label{fig9}}
 
Next we will investigate  the impact of the neutrino velocities
assigned in the initial conditions.  In Figure \ref{fig7} we show 
for our default simulations with  $z_{\rm  IC}=7$, the ratio of the
matter power spectrum for a simulation without the thermal
velocities relative to the matter power spectrum of a simulation where
the velocities have been included in the initial conditions. Without
the velocities  in the initial conditions the power is less suppressed 
at $k\sim 1\,h/$Mpc by roughly 3\%, 2\%
and 1\% at $z=2$, $3$, and $4$, respectively, with a very weak dependence on the
wavenumber considered. These values are in good agreement with the
$z=0$ results reported by Ref.~\cite{brandbyge08}.

Another effect that could potentially affect our results are  the  details
of the sampling of the initial phase-space density distribution of the neutrinos.
Ref.~\cite{maber94} suggested that it would be advantageous to conserve
momentum by creating pairs of neutrinos with equal and opposite thermal
velocities.  To test this, we modified the initial condition code  to
produce neutrino pairs with mass $m_p=m_{\nu,p}/2$, instead of a single
neutrino particle with mass $m_{\nu,p}$. The two
neutrino particles are then assigned the same velocities in opposite
directions in order to conserve momentum. The results are shown in Figure
\ref{fig8}, where we report the ratios of neutrino power spectra in the left
panel and that of the  matter power spectra in the right panel. The impact is 
very small  and is fully accounted for by the different number of neutrino
particles used, which decreases the Poisson contribution in the  case
of neutrino pairs by a factor of two at $k> 2-3\,h/$Mpc.

The last effect that we examine here is the dependence of the matter
power spectrum on the initial redshift of the simulation
\cite{wangwhite,gaotheuns}.  For this purpose we have performed four
additional simulations with initial redshifts $z=4$ and $z=49$ for the
simulation with \smnu = 0.6 eV and the $\Lambda$CDM simulation without
neutrinos. In Figure \ref{fig9}, we plot the suppression due to the
effect of neutrinos for the matter power spectrum at $z=2$ and $3$
(black and red curves) for the three different values of the starting
redshift. At $k\sim 3h/$Mpc there are differences of the order of 10\%
(3\%) at $z=3$ ($z=2$). As expected the results for our default
simulation lie between those for the low and the high starting
redshift. For the early starting redshift ($z=49$) the neutrino
component becomes effectively Poissonian even at the largest scales,
since the neutrino power spectrum as computed by {\small{CAMB}} will
be in general much smaller at high redshift than the Poisson
contribution (which is independent of redshift): this translates into
a larger overall suppression of the matter power spectrum at small
scales.  Note that the different amount of Poisson power with respect to
 physical neutrino clustering in the initial conditions has an impact
 also on the subsequent clustering of the neutrino and matter
  components.  For the low starting redshift ($z=4$), the relevant
scales are already affected by mildly non-linear growth.  Note further that
the relatively strong dependence on the initial redshift at $z=2-4$ is
much more pronounced than that inferred by
Refs.~\cite{brandbyge08,brandbyge09a}, who found an overall agreement
at the percent level at $z=0$ between simulations with different
initial redshifts. This can probably be attributed at least partially
to their use of second-order Lagrangian corrections in the initial
conditions, which reduces the errors introduced when a low starting
redshift is used. In addition, these authors studied much larger
scales and by $z=0$ the non-linear evolution tends to largely erase
the memory of differences in the initial conditions on small scales.
For the purposes of our study $z=7$ appears to be an acceptable
compromise.  We provide further support for this in the next sections
where we quantify the impact of the starting redshift on the \lya flux
power spectrum.

\section{The effect of neutrinos on statistics of the flux  distribution in the \lya forest }

\subsection{Flux power and flux probability distribution function}

In this Section we focus on  the effect of neutrinos on the matter
distribution as probed by the IGM,
and in particular on the transmitted \lya flux and its one-point flux
probability distribution function (PDF), and its two-point statistics (the
flux power spectrum).  To perform our analysis we have extracted 1000 mock quasar
absorption spectra from the simulations  at many different
redshifts.  All spectra are constructed in redshift space, taking into account
the effect of the peculiar velocities of the IGM $v_{\rm pec,\parallel}$ along the
line-of-sight. The flux at  redshift-space coordinate
$u$ (in km/s/Mpc) can be written as $F(u)=\exp[-\tau(u)]$ with
\begin{equation}
\tau(u)={\sigma_{0,\alpha} ~c\over H(z)} \int_{-\infty}^{\infty} dx\, n_{\rm  HI}(x) ~{\cal G}\left[u-x-v_{\rm pec,\parallel}^{\rm   IGM}(x),\,b(x)\right] {\rm d}x \label{eq1} \;, 
\end{equation}
where $\sigma_{0,\alpha} =4.45 \times 10^{-18}$ cm$^2$ is the hydrogen \lya cross-section, $H(z)$ is the
Hubble constant at redshift $z$, $x$ is the real-space coordinate (in km
s$^{-1}$), $b=(2k_BT/mc^2)^{1/2}$ is the velocity dispersion in units of $c$,
${\cal G}=(\sqrt{\pi} b)^{-1}\exp[-(u-y-v_{\rm pec,\parallel}^{\rm
    IGM}(y))^2/b^2]$ is a  Gaussian profile that  approximates the Voigt
profile well in the regime considered here. 

The neutral hydrogen density in real-space in  the equation above is
approximately related to the underlying gas density
(e.g. \cite{huignedin97}) as, 
\begin{equation} 
n_{\rm HI}({\bf r}, z) \approx 10^{-5} ~{\overline n}_{\rm IGM}(z)
\left({\Omega_{0b} h^2 \over 0.019}\right) \left({\Gamma_{-12} \over
  0.5}\right)^{-1} \times \nonumber \left(T({\bf r},z) \over 10^4 {\rm
  K} \right)^{-0.7} \left({1+z \over 4}\right)^3 \left(1 + \delta_{\rm
  IGM}({\bf r},z) \right)^2 \;,
\label{eqneu}
\end{equation}
where $\Gamma_{-12}$ is the hydrogen photoionization rate in units of
$s^{-1}$, $T$ is the IGM temperature, $\overline{n}_{\rm IGM}(z)$ is
the mean IGM density as a function of redshift and ${\bf r}$ is the
real-space coordinate.

As we have the benefit of a full hydrodynamical simulations there
is, however, no need to make the approximations  underlying
equation~(\ref{eqneu}). We calculate the integral in eq.~(\ref{eq1})
to obtain the \lya optical depth along each simulated line-of-sight 
using directly the relevant hydrodynamical quantities from the
numerical simulations: $\delta_{\rm IGM}, T, v_{\rm pec}, n_{\rm HI}$. Further
details on how to extract a mock quasar spectrum from an hydrodynamical
simulation using the SPH formalism can be found in \cite{theuns98}. We 
have added noise typical for observed spectra and convolved the spectra with 
the instrumental resolution corresponding to observed high-resolution
spectra. Note that the resolution has a larger effect on the  flux PDF 
than the flux power spectrum. The ensemble  of all our spectra have
then been normalized by adjusting the assumed Ultra Violet background such that
the observed mean flux level in high-resolution quasar spectra \cite{kim07} at a given redshift,
$<F(z)>=\exp(-\tau_{\rm eff}(z))$ with $\tau_{\rm eff}(z) = 0.0023 \,
(1+z)^{3.65}$ is reproduced.

\FIGURE
{\includegraphics[width=15.5cm]{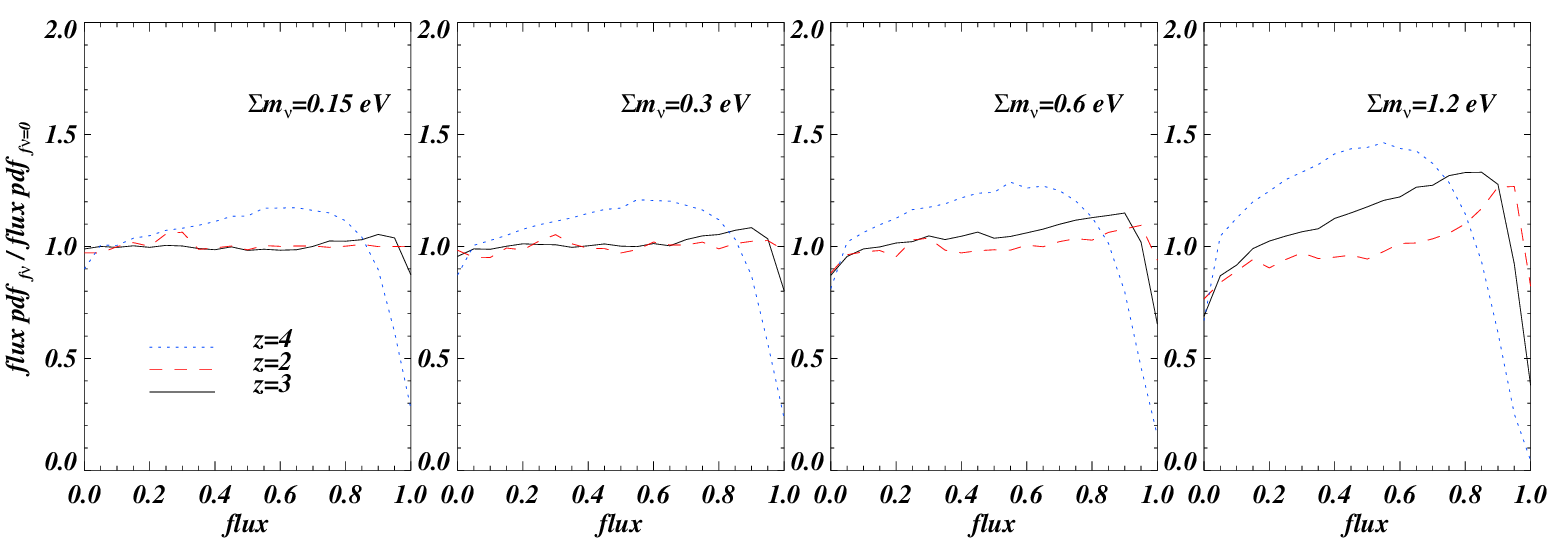}
\caption{{\it Effect of $f_{\nu}$ on the flux probability distribution
    function} for four different neutrino masses, $\Sigma m_\nu =0.15$, 0.3, 0.6, $1.2$~eV,
  from left to right. Different line-styles  refer to different
  redshifts: $z=2$ (red dashed), $3$ (black continuous) 
  and $4$ (blue dotted).}
   \label{fig10}}

The \lya flux PDF is very well measured especially from high
resolution quasar spectra (the statistical errors are at the percent
level).  We recently performed a careful analysis of the systematic
uncertainties and were able to extract interesting astrophysical and
cosmological constraints from the flux PDF \cite{pdflya}.  In
Figure~\ref{fig10}, we show the ratio between the flux probability
distribution functions of simulations with and without neutrinos: from
left to right, the cases for \smnu$=0.15$, $0.3$, $0.6$, and $1.2$~eV
are reported at $z=2$, $3$, and $4$. The larger the neutrino masses
\smnu, the more peaked the flux distribution becomes at intermediate
flux values. The reason for this trend is that the growth of structure
is suppressed in the simulation with neutrinos. As a result, voids
(flux$\sim 1$) are less empty and clustered regions are less dense
than in the simulation without neutrinos and the effect is stronger at
high redshift than at low redshift.

\FIGURE
{\includegraphics[width=15.5cm]{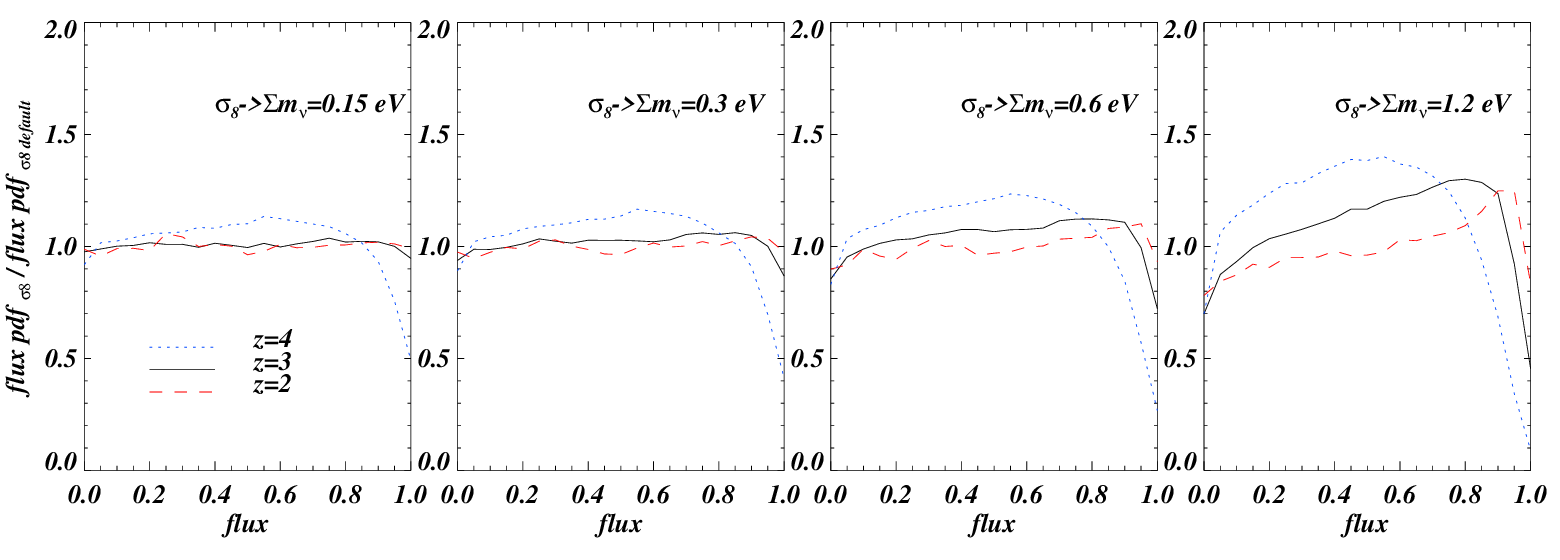}
\caption{{\it Effect of different r.m.s. values for the amplitude of the
    matter power on the flux probability distribution}. Four cases are
  presented that have exactly the same $\sigma_8$ at $z=7$ as those of the models
  $\Sigma m_\nu =0.15$, 0.3, 0.6, $1.2$~eV, from left to right. Different
  line-styles  refer to different redshifts:  $z=2$ (red dashed), $3$ (black continuous) 
  and $4$ (blue dotted).}
   \label{fig11}}

Analogous to our discussion in the previous section we also compute
the flux properties for simulations without neutrinos but with a
reduced overall amplitude of the matter power spectrum normalized 
to the same $\sigma_8$ at $z=7$. The results are shown in Figure
\ref{fig11}.  The trends with neutrino mass are similar to those seen in Figure \ref{fig10}, 
but slightly less  pronounced.

\FIGURE
{\includegraphics[width=15.5cm]{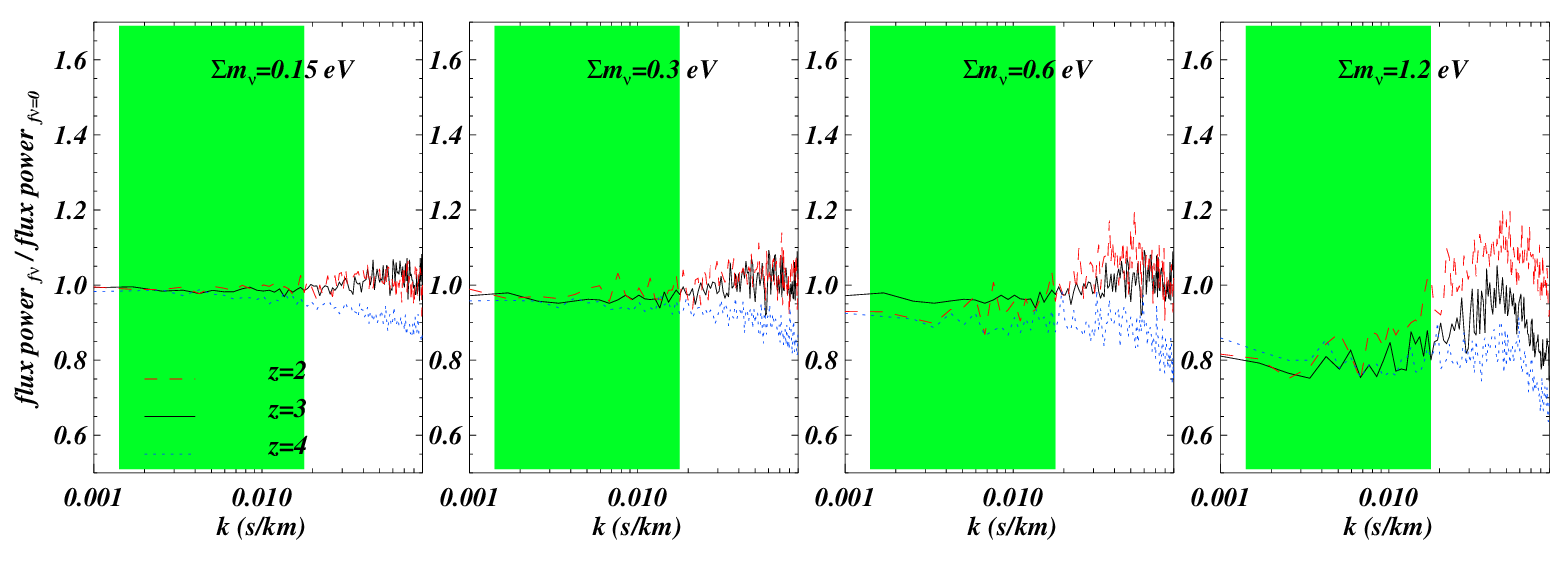}
\caption{{\it Effect of $f_{\nu}$ on the flux power spectrum}. Ratio
  between flux power spectra with and without neutrinos as a function
  of wavenumber in s/km. Four different cases are presented with
  $\Sigma m_\nu=0.15$, 0.3, 0.6, $1.2$~eV, from left to
  right. Different line-styles refer to different redshifts: $z=2$
  (red dashed), $z=3$ (black continuous), and $z=4$ (blue dotted). The
  shaded area indicates the range of wavenumbers probed by the SDSS
  flux power spectrum.}
   \label{fig12}}


We now turn to the flux power spectrum, a quantity which is more
closely related to the underlying matter power and has been
extensively used to constrain cosmological and astrophysical
parameters (e.g. ~\cite{croft02}). The \lya flux power spectrum
provides a more direct link to the matter power spectrum than the flux
PDF: it is sensitive to cosmological parameters, the thermal state of
the IGM, instrumental effects (signal to noise and resolution), the
presence of metal lines and the nature of dark matter at small scales,
etc. (see for example \cite{kim04,mcdonald05}). The flux power spectrum  has been measured 
over a wide redshift range,  $z=2-5.5$,
using both high and low-resolution data. The growth of cosmic
structures can thus be constrained over a significant fraction of the
cosmic time, lifting the degeneracies between astrophysical and
cosmological parameters that present different redshift and scale
dependencies in this range.

We show the measured flux power spectra for our different simulations
in Figures \ref{fig12} and \ref{fig13}.  Note that the results have
not been smoothed. We recall that the useful range of high resolution
spectra reaches to $k=0.03$ s/km while we can reach to $k\sim$ 0.018
s/km with low-resolution SDSS spectra. We are here primarily
interested in quantifying the effects over this range of wavenumbers.

For \smnu=0.15~eV, the only effect of neutrinos on the flux power is a
$<5\%$ suppression at $z=4$. As expected the effect becomes larger
with increasing neutrino mass. At the largest scales the flux power in
the simulations with neutrinos is suppressed by 5, 7 and 15\% for
\smnu = 0.3, 0.6 and 1.2 eV, respectively. There is some dependence of
the suppression on wavenumber with an upturn at small scales of about
0.01 s/km and a bump at $k\sim 0.05$~s/km.

The relationship between one-dimensional flux power spectrum and
three-dimensional matter power is non-trivial, not only because of the
fact that the one-dimensional matter power is an integral of the
three-dimensional spectrum, but also due to non-linearities in the
flux-density relation. As clearly demonstrated in
Ref.~\cite{vielcarswell}, systems with column densities $\sim 10^{14}$
cm$^{-2}$ contribute most to the flux power at $k\sim 0.05$ s/km, and
these absorbers are produced by gas which is close to the mean density
\cite{schaye}. The differences in the flux power spectrum of
simulations with and without neutrinos reflect the differences in the
spatial distribution of gas in models which  have experienced
different amounts of growth of structure: at $z<3$, a model with a
reduced amplitude of the matter power spectrum has more structure at
mean density than a high-$\sigma_8$ model for which the gas
probability distribution function is more skewed. Note that the
suppressions for the simulations without neutrinos in Figure
\ref{fig13} are very similar to those with a reduced amplitude of the
matter power spectrum with the same value of $\sigma_8$. The small
differences visible in the three dimensional matter power spectra are
thus even smaller in the flux power spectrum.

\FIGURE
{\includegraphics[width=15.5cm]{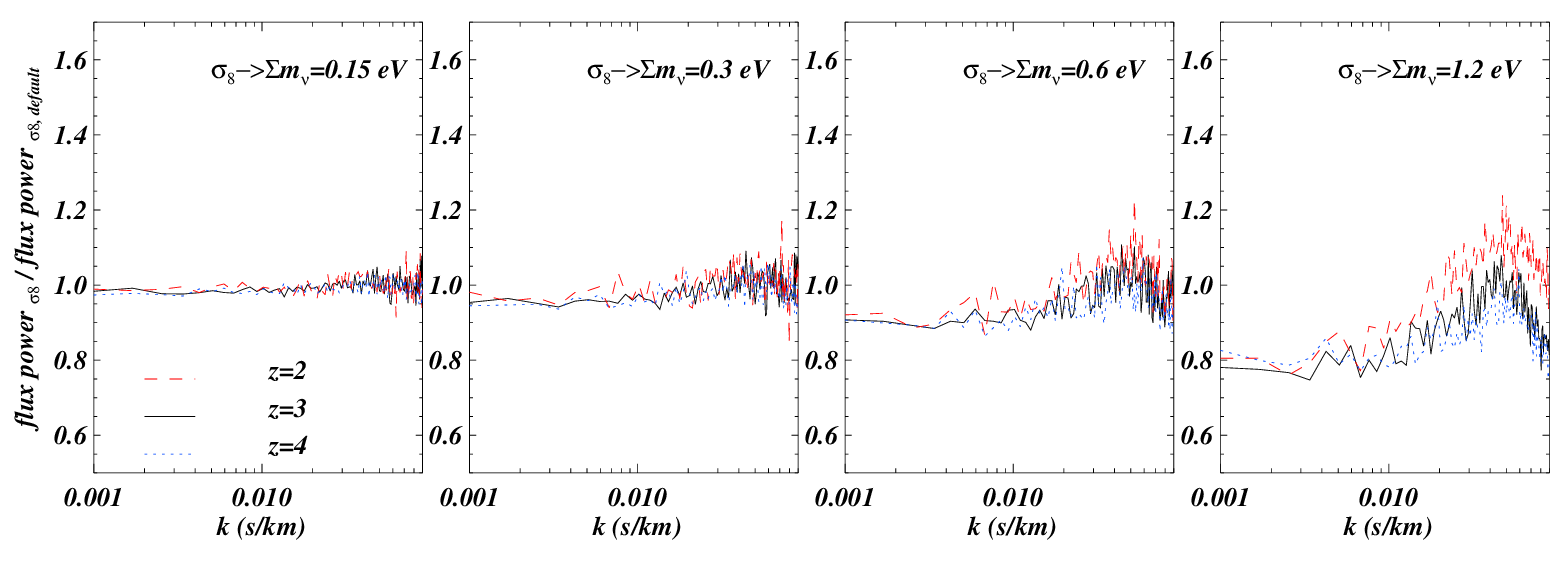}
\caption{{\it Effect of different r.m.s. power spectrum amplitudes on the
    flux power spectrum}. Four different cases are presented that have exactly the same
  $\sigma_8$ at $z=7$ as those of the models $\Sigma m_\nu =0.15$, 0.3, 0.6, $1.2$ eV,
  from left to right. Different line-styles refer to different
  redshifts:  $z=2$ (red dashed), $z=3$
  (black continuous), and $z=4$ (blue dotted).}
   \label{fig13}}


\subsection{Numerical effects on the flux power spectrum for simulations with the particle based implementation of neutrinos}

In this subsection we  explore the sensitivity of our results for the
flux power spectrum  on a number of numerical effects.  In Figure~\ref{fig14}, we
show the ratios of the flux power spectrum for simulations with  \smnu=0.15 eV (left
panel) and \smnu=0.6 eV (right panel) with   $N_{\nu}=512^3$ and
$N_{\nu}=1024^3$ neutrino particles. There is an  opposite trend here 
to  what we found in the corresponding plot for the matter power shown in
Figure \ref{fig6}. There is a  bump at $k\sim 0.05$ s/km where the
ratio rises above unity. The ratio of the  matter power spectrum 
at a similar scale of $5 h/$Mpc does not
change or is mildly suppressed. At these scales the matter power
spectrum is affected by Poisson noise due to the neutrinos. 
The larger  Poisson contribution to the matter power spectrum
in the simulation with the smaller  number of neutrinos   should result in
less diffuse small scale absorbers and lower  the amplitude of the
flux power spectrum at scales $>0.01$ s/km. 

We have also checked the effect of a different number of mesh points
on the PM grid by running a simulation with $N_{\nu}=1024^3$ and a PM
grid of $512^3$ mesh points. We find that the impact is negligible in
the range of scales of interest, about $\pm 5\%$ ($\pm 1\%$) at scales
smaller than $k>0.2 \,(0.06)$ s/km. Note however that these scales are
much smaller than those we are interested in.

\FIGURE
    {\includegraphics[width=14cm]{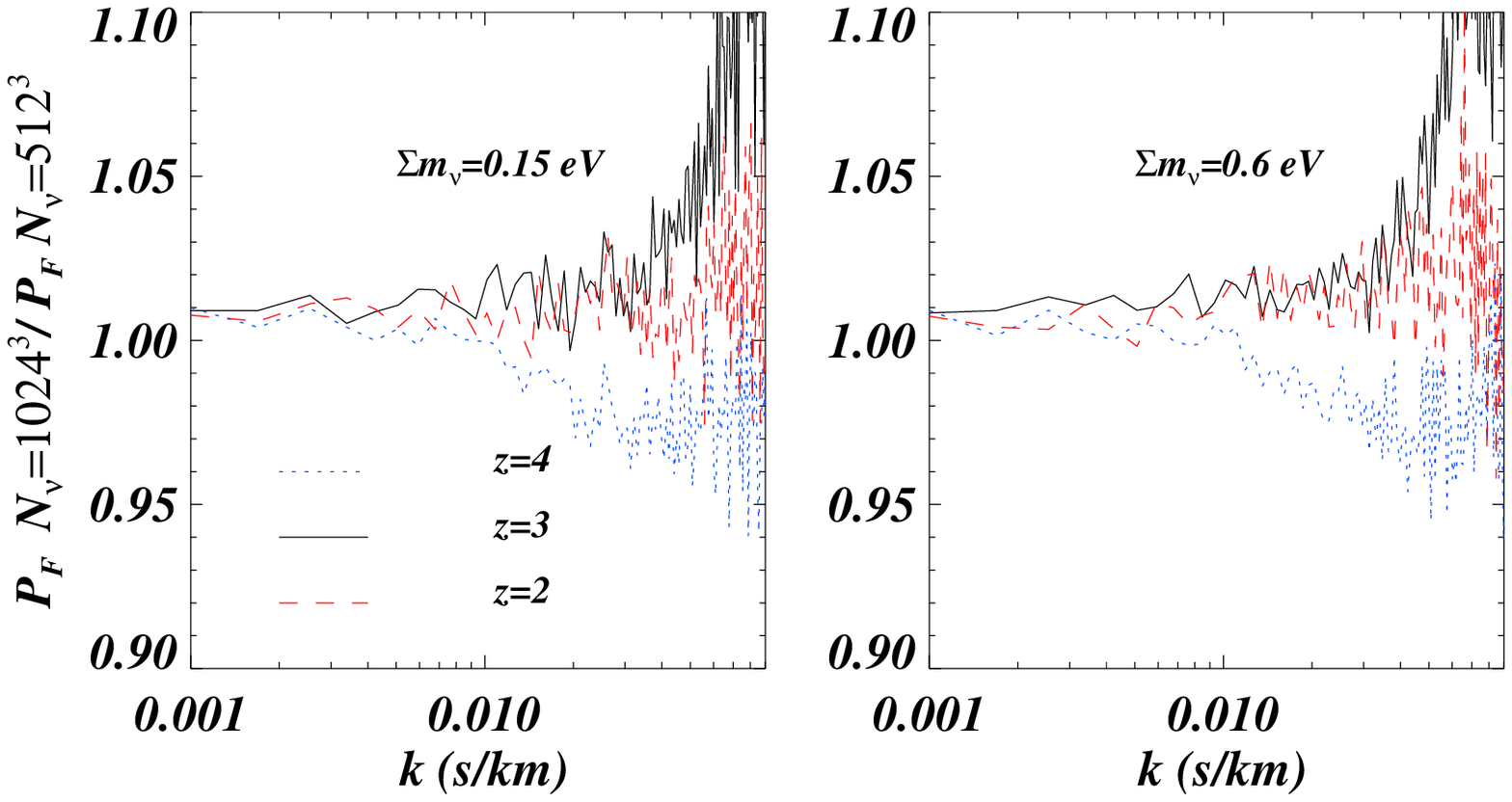}
\caption{{\it Effect of different number of neutrino particles on the flux
    power}. Comparison of simulations with  $N_{\nu}=512^3$ and $N_{\nu}=1024^3$
    neutrino particles with  masses of  $\Sigma m_{\nu} =0.15$ eV (left panel)
  and $\Sigma m_{\nu} =0.6$ eV (right panel), at $z=2$, $3$, and
  $4$.}\label{fig14}}

\FIGURE
{\includegraphics[width=9cm]{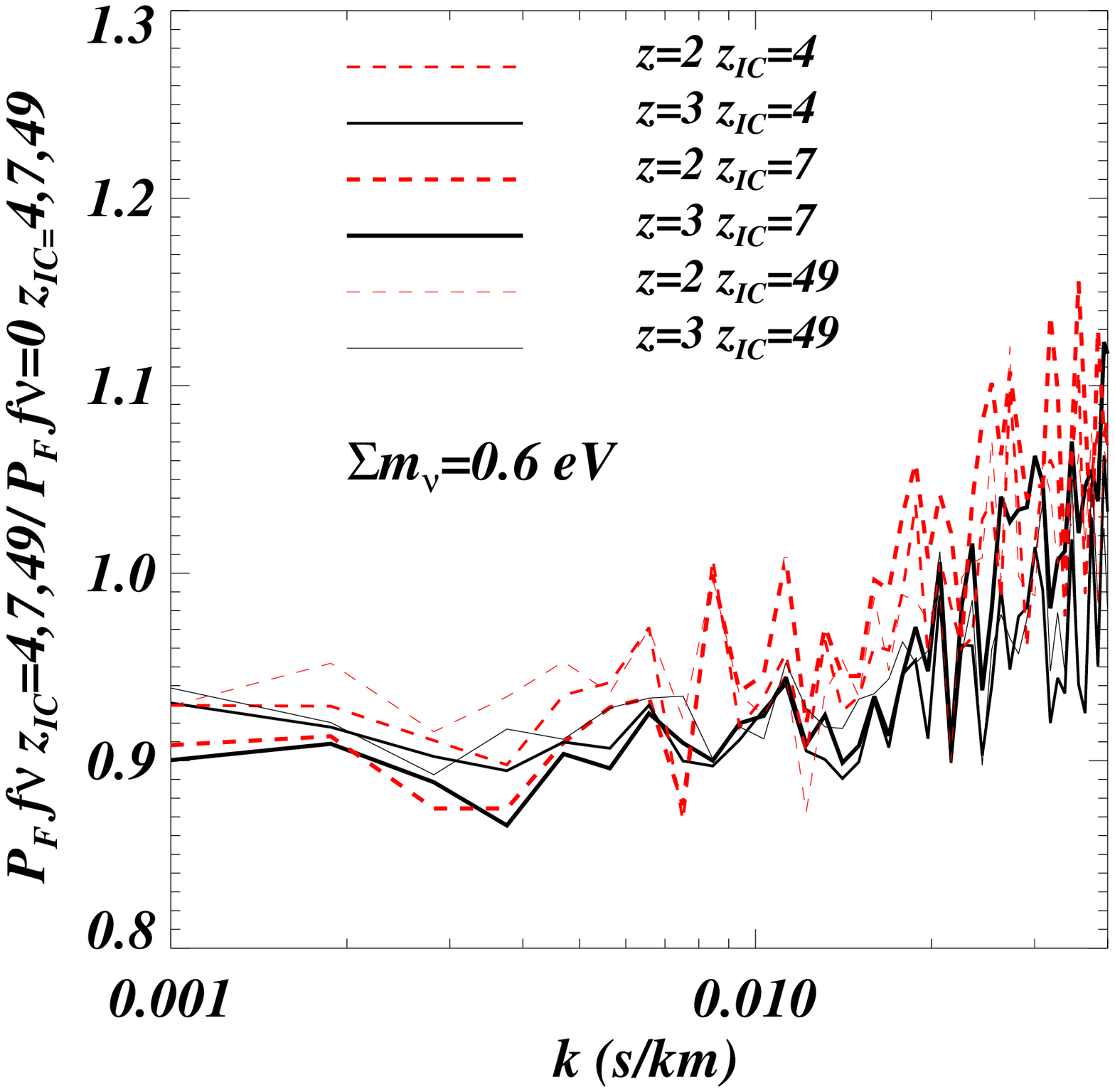}
\caption{{\it Effect of different initial redshifts on the flux power
    spectrum }. We show
  the following quantities at $z=2$ (red dashed) and $z=3$ (black continuous):
  $P(k,f_{\nu},z_{IC}=4)/P(k,f_{\nu}=0,z_{IC}=4)$ (thin curves),
  $P(k,f_{\nu},z_{IC}=7)/P(k,f_{\nu}=0,z_{IC}=7)$ (very thick curves and
  default case) $P(k,f_{\nu},z_{IC}=49)/P(k,f_{\nu}=0,z_{IC}=49)$ (thick
  curves). All simulations shown are for  $\Sigma m_{\nu} =0.6$ eV.}\label{fig15}}

In Figure \ref{fig15} we show the effect of varying the starting
redshifts on the flux power spectrum ratios. The differences are
significantly smaller than those for  the matter power spectra  
(Figure~\ref{fig9}).  For neutrino masses 
with  \smnu=0.6 eV the differences at $z=2$ and $z=3$ are at the
level of 2\% or less over the whole range of relevant wavenumbers 
(note that curves of the same color should be compared with each other).

The flux power spectrum of our simulation at $z=2-4$ appears to have  
converged at the 2\% level in the range $k \in [0.001, 0.03]\,$ s/km.  This
level of numerical convergence should be sufficient for the analysis of
presently available  high- or low-resolution \lya forest data. The
SDSS \lya flux power spectrum has statistical errors of the order of
3\% or more at the smallest scales, and up to 10-15\% at the largest
scales, while the statistical errors of  the high-resolution data  are about two times larger than these.

\subsection{Grid based {\it vs} particle based simulations of the effect of neutrinos on \lya forest data}

We now have a closer look at the relative merits of grid based and
particle based simulation on scales relevant for \lya forest data
especially with regard to the  \lya flux power spectrum.  We will show how the
differences  shown in Figure \ref{figcompare1} for the matter power spectra
propagate to the flux power spectrum.
In the left panel of Figure \ref{figgrid} we show the suppression of the matter power spectrum due
to the free-streaming of neutrinos with \smnu=0.6 eV model on the
matter power for simulations with the grid based implementation of the
neutrino density.  The suppression is larger than that of the
corresponding simulation with the particle based neutrino
implementation by about 10\% (the horizontal green thick line
indicates a value of $-12\, f_{\rm \nu}$).  In the middle panel, we
directly compare the two implementations and while it is evident that
at still linear scales, $k<0.8 h$/Mpc, the agreement is at the 2\%
level, at smaller scales the differences are larger, about 7\% at
$k\sim k_{\rm max}$.  In the rightmost panel, we compare the two
implementations in terms of (one-dimensional) flux power.  In this
case the differences are of the order of $<4\%$ for the scales
considered in here.  On the scales relevant for the \lya forest data
the non-linear evolution of the matter distribution is more important
than the effect of the Poisson contribution to the neutrino power
spectrum justifying our choice of the particle based implementation
for a quantitative analysis.

\FIGURE
{\includegraphics[width=15.5cm]{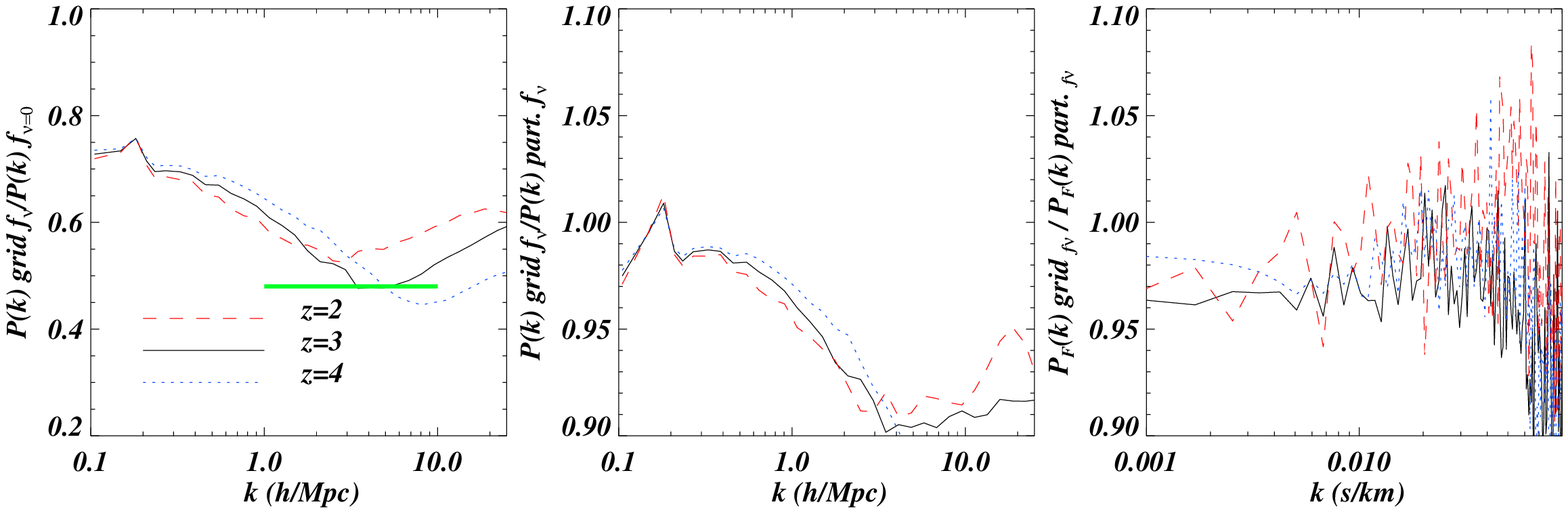}
\caption{{\it Effect of neutrinos on the matter and flux power for a
    grid based implementation}.  Suppression induced by a \smnu=0.6 eV
  model (left panel) at $z=2,3,4$ (dashed red, continuous black,
  dotted blue curves, respectively in the right panel). An overall
  suppression of $-12 f_{\rm \nu}$ is shown as a thick green
  line. Ratio between the grid and the particle based implementations
  (middle panel). Impact on the flux power (right
  panel).\label{figgrid}}}

\section{An upper limit on the neutrino mass from the SDSS \lya forest data}

\subsection{The SDSS flux power spectrum}

We now turn to deriving  an upper limit on the neutrino mass 
from the SDSS \lya forest data  for which  the flux
power spectrum has been measured by \cite{mcdonald06}. This unique data set consists
of 3035 absorption spectra of quasars in the redshift range $2<z<4$, drawn from the
DR1 and DR2 data releases.  Since the spectral resolution is $R\sim 2000$, the
typical \lya absorption features with a width of $\sim 30$ km/s are not
resolved.  The signal-to-noise of the individual spectra is rather low, S/N
$\sim 5$ per pixel. Due to the  large number of spectra the flux
power spectrum on scales a factor of a few larger than the thermal cut-off can,
however, be measured with small statistical errors.  

Ref.~\cite{mcdonald06} have re-analyzed the raw data and have
extensively investigated the effect of noise, resolution of the spectrograph, sky
subtraction, quasar continuum and associated metal absorption: corrections for
these effects are made and estimates of the associated errors are given. The
noise contribution to the flux power spectrum rises from 15-30 percent at the
smallest wavenumbers to order unity at the largest wavenumbers and varies with
redshift.  The correctios  for uncorrelated  metal absorption are generally a
factor five to ten smaller than this. The correction for resolution varies
from 1\% at the smallest wavenumbers to a factor four at the largest
wavenumbers. As final result of their analysis, \cite{mcdonald06} present
their estimate for the flux power spectrum $P_F(k,z)$ at 12 wavenumbers in the
range $0.00141<k\,$(s/km)$<0.01778$, equally spaced in $\Delta \log k = 0.1$
for $z=2.2$, 2.4, 2.6, 2.8, 3.0, 3.2, 3.4, 3.6, 3.8, 4.0, and 4.2, for a total
of 132 data points. They also provide  the covariance matrix. Here, we
will use this flux power spectrum together with the recommended corrections to
the data and the recommended treatment of the errors of these corrections.

We note that Ref.~\cite{mcdonald05} has interpreted this flux power spectrum
previously based on a set of numerical simulations giving a measurement of the
linear matter power spectrum at $z=3$ and $k=0.009$ s/km. Due to the wide
redshift range sampled, many degeneracies between cosmological and
astrophysical parameters can be broken, allowing for a high precision
measurement of the linear power spectrum at small scales.

\subsection{Multi-dimensional likelihood analysis}
In order to explore the multi-dimensional astrophysical and cosmological
parameter space we will use an improved version of a method based on 
a Taylor expansion of the flux power spectrum around a fiducial
model  as presented in Ref.~\cite{vh06}. Note that this is an approximate
approach that assumes a physically motivated best-guess model and allows an
exploration of the likelihood function around it.  If we denote with ${\bf p}$
an arbitrary vector of astrophysical, cosmological and noise-related
parameters close to the best-guess model described by ${\bf p_0}$, we assume
that:
\begin{equation}
P_F(k,z;{\bf p})=P_F(k,z;{\bf p^0})+\sum_i^N \frac{\partial {P_F(k,z;p_i)}}{\partial {p_i}}\bigg|_{\bf
{p}=\bf{p^0}}(p_i - p_i^0)+\sum_i^N \frac{\partial^2 {P_F(k,z;p_i)}}{\partial {p_i}^2}\bigg|_{\bf
{p}=\bf{p^0}}\frac{(p_i - p_i^0)^2}{2}\, ,
\label{taylor}
\end{equation}
where $p_i$ are the $N$ components of the vector ${\bf p}$, which represent
the astrophysical and cosmological parameters. We perform here
the Taylor expansion to second order for each parameter independently
(i.e. neglecting cross derivatives).  To obtain the derivatives of the
flux power spectrum we  run a suite of  hydrodynamical simulations changing  one
parameter at a time with respect to those of the best-guess model
and keeping all other parameters fixed. We then calculate the first and second order
coefficients according to equation~(\ref{taylor}).  This procedure is
performed for the astrophysical parameters describing the thermal state of the
IGM, $T$ and $\gamma$ (both described by a power-law at $z=3$ with three
parameters each: amplitude at $z=3$, and power-law indices for $z<3$ and
$z>3$), the mean flux level (amplitude and slope) and the reionization
redshift; and for the cosmological parameters in a flat $\Lambda$ CDM model ($n_s$,
$\sigma_8$, $H_0$, $\Omega_{\rm 0m}$, \smnu) that affect the growth of structure.

We then use a Monte-Carlo Markov Chain technique based on a suitable modification
of the publicly available {\small COSMOMC} code \cite{cosmomc}. We
thereby  allow for parameters  that describe noise, resolution and those that model
the contribution of Damped \lya systems to the flux power spectrum.  In total a
set of 28 parameters  are allowed to vary. For a more extensive discussion
on the use of the  MCMC method in this context  we refer to \cite{boya09,v09,lidz09}.

\FIGURE
{\includegraphics[width=11cm]{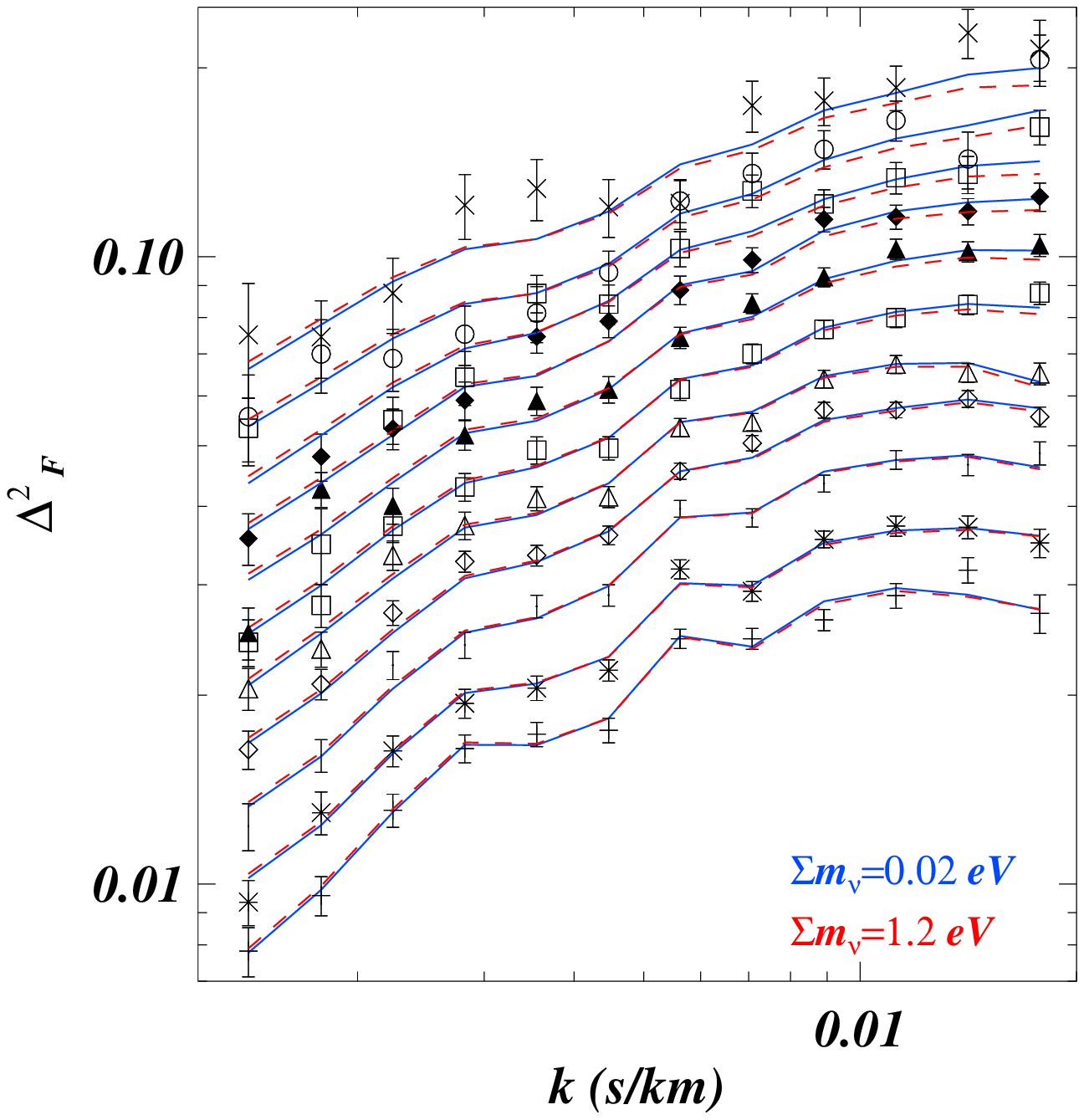}
\caption{{\it Effect of different values of \smnu on the flux power
    spectrum for simulations normalized at \lya forest scales}. 
    The solid  blue curves show the the best-fit model to
    the SDSS flux power spectrum (data points with error bars), a model
    which is ruled out by the data at $>2 \sigma$ level is also reported
    (red dashed curves).\label{figsdss}}}

\FIGURE
{\includegraphics[width=11cm,height=6cm]{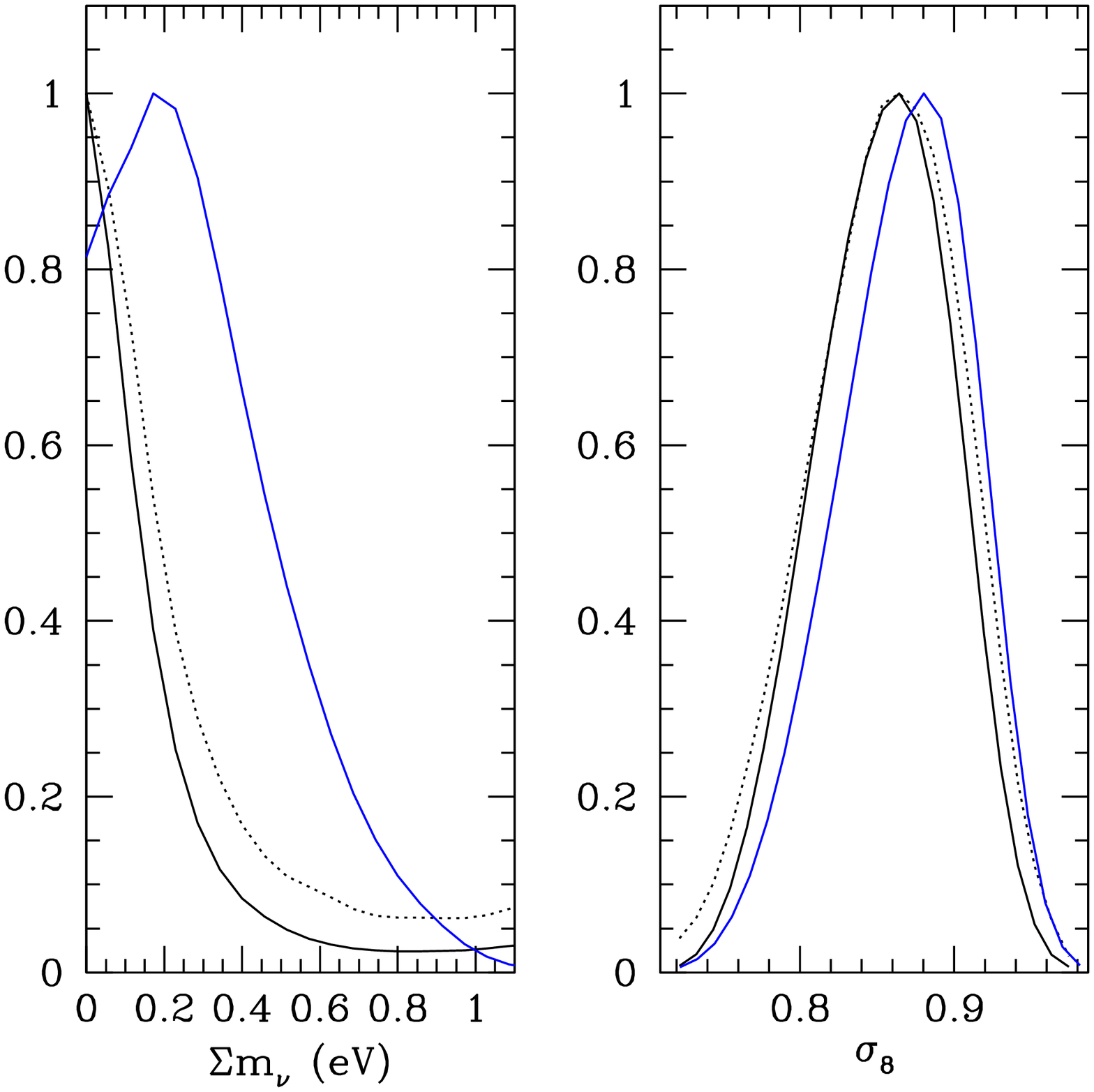}
\caption{{\it One-dimensional mean and marginalized likelihoods for
    the r.m.s. value of the matter power spectrum amplitude and
    \smnu}. The likelihoods computed from the Monte-Carlo Markov
  Chains from the SDSS flux power spectrum using the simulations
  including neutrinos shown for \smnu (left panel) and
  $\sigma_8$ (right panel). The blue curves represent the results
  obtained if the effect of neutrinos is approximated by changing 
  $\sigma_8$ in simulations without neutrionos. 
   Mean  likelihoods are represented by the dotted curves, while 
  marginalized likelihoods  are shown as continuous curves. \label{figlike}}}

The main difference to our previous work in this respect is the addition
of the effect of the neutrino mass \smnu on the flux power
spectrum.  Our analysis here does thereby not rely on any additional
data  which independently constrains the
amplitude of the matter power spectrum on large scales.
The results obtained in the previous sections refer to
simulations that have a different $\sigma_8$ and $\Omega_{\rm 0m}$ than our
best-guess model ($\sigma_8=0.85, \Omega_{0m}=0.26, n_S=0.95$).
We have thus rescaled the flux power spectrum  to adjust for the different
value of $\sigma_8$, $P_F(k,z) [f_{\nu},\sigma_8=0.85]=
P_F(k,z)[f_{\nu},\sigma_8]\times
P_F(k,z)[f_{\nu}=0,\sigma_8=0.85]/P_F(k,z)[f_{\nu}=0,\sigma_8]$.

The effect of the free-streaming of the neutrinos  \smnu $> 0$ for a
fixed value of $\sigma_8$ is a small additional scale dependent suppression of the 
flux power which depends on redshift and mass of the neutrinos. The
difference between  the thick and thin curves in  Figure~\ref{fig4} 
shows the effect for the matter power spectrum. 
Comparison of  Figure~\ref{fig12} with Figure~\ref{fig13}
shows that the corresponding effect on the flux power spectrum 
is somewhat smaller.  

In order to demonstrate  the effect of different \smnu values  on the
flux power spectrum we compare the theoretical predictions directly with the
SDSS data points in Figure ~\ref{figsdss}. Two different
models are shown: the best fit model to the data with \smnu=0.02 eV and the
same model (i.e. all the parameters fixed to the same values) but with
a different \smnu=1.2 eV. The constraining power is largest  for the
data points at small scale and high redshift, similar to studies 
which constrain the free-streaming by  warm dark matter particles.

Our results with regard to the upper limit on the mass of neutrinos
are summarized in Figure ~\ref{figlike}, where the one-dimensional
marginalized (continuous curves) and mean (dashed curves) likelihoods
for \smnu and $\sigma_8$ are shown in the left and right panels,
respectively. The constraints are $\sigma_8=0.85\pm0.04$ (1$\sigma$
error bars) and \smnu$<0.86$ eV (2$\sigma$ confidence level). The
value for $\sigma_8$ is in good agreement with previous analyses
(e.g.~\cite{vh06}). We stress again that the constraint on \smnu is
obtained from the SDSS flux power spectrum {\em alone} without
considering other external data sets.  The $\chi^2$-value for the best
fit model is 138.3 for 129 degrees of freedom, which should occur with
a probability of 11\%. We regard this upper limit as a conservative
one, since having used the results obtained with the grid based
approach would have produced a lower value than the one presented
here, being the neutrino induced suppression larger for grid based
simulations.

We note that the likelihoods in Figure~\ref{figlike} stay somewhat
flat even for \smnu values $\sim 1$~eV, meaning that these values are
allowed by SDSS data at the 2-2.5 $\sigma$ level. In this Figure the
mean likelihood is represented by the dotted curve, while the
marginalized one is shown as a continuous curve.  Note that this is
very different from the result of the analysis in \cite{seljak06},
where a very low upper limit of \smnu=0.17~eV was obtained. There are,
however a number of important differences to our study here. The study
of \cite{seljak06}, is based on: $i)$ inferring the linear dark matter
power spectrum with a suite of approximate hydrodynamical simulations
that do not incorporate neutrinos; and $ii)$ combining this
measurement with other large scale structure probes. As extensively
discussed in \cite{seljak06} the tension between the high
r.m.s. values for the amplitude of the matter power spectrum suggested
by \lya data and the lower values inferred from cosmic microwave
background experiments is the main reason for the very low limit on
\smnu.

It is also interesting to compare our findings with those that would
be obtained by not considering the additional non-linear effects in
simulations including neutrinos on the {\sl shape} of the \lya flux
power spectrum.  The effect on the flux power spectrum can in this
case be captured by the degenerate effect of changing the value of
$\sigma_8$.  In order to investigate this, we first add an extra
parameter to represent neutrino's mass fraction in the markov chains,
which has exactly the same flux derivatives of the corresponding
$\sigma_8$ value, and then vary it independently from the others
obtaining \smnu$<0.75$ eV ($2\sigma$ C.L).  The results are shown in
Figure \ref{figlike} and one can see that the hold method results in a
tighter upper bound because the likelihood is less flat for large
\smnu values than in the new method (the $1\sigma$ upper limits are
0.38 eV and 0.18 eV for the old and new method, respectively).  With
this approximation the constraints are thus only slightly tighter than
if we use the results from simulations including neutrinos which take
the full non-linear effects on the shape of the flux power spectrum
into account. The upper limit obtained in this way is also comparable
to the constraint derived by just mapping (a-posteriori) the $2\sigma$
lower limit on $\sigma_8 \sim 0.75$, into an upper limit on \smnu
using the code {\small{CAMB}}.  In this latter case, however,
$\sigma_8$ and \smnu are not treated as independent parameters as they
should if the matter power spectrum is normalized at the \lya forest
scales. We think that the somewhat tigther constraints obtained
without the results from the numerical simulations including neutrinos
is due to the method we used. With the Taylor-expansion method we
model small departures from a best-guess case and this is more
accurately described by implementing the exact results on the flux
power from the numerical simulations rather than treating the effect
of \smnu and $\sigma_8$ independently (in this second case the flux
power is less likely to depart significantly from the reference case
than in the first case). However, given the small difference between
the two methods we would not regard this discrepancy as particularly
significant.

We have focused here instead on investigating the impact  of neutrinos 
on the (non-linear) spatial distribution of the neutral hydrogen in the IGM 
and the resulting flux power spectrum and obtaining a consistent upper
limit on neutrino masses from the SDSS flux power spectrum alone. 
The rather small but scale dependent and redshift dependent impact 
of neutrinos with \smnu$=0.9-1$ eV at the 4-5\% level already results 
in an interesting upper limit which is somewhat stronger than that
from the CMB data alone.

\section{Summary and Discussion}

\lya forest data in combination with cosmic microwave background
measurements provide presently the lowest upper limits on the masses
of neutrinos. A careful assessment of the ability of \lya forest data
and in particular the \lya flux power spectrum to put limits on the
effect of the free-streaming of neutrinos on the matter distribution
is therefore important. The use of \lya forest data for measurements
of the matter power spectrum relies heavily on the accurate modeling
of the spatial distribution of neutral hydrogen.  We have presented
here for the first time a study of a large suite of hydrodynamical
cosmological simulations that allow a quantification of the impact of
neutrinos on the non-linear matter distribution as probed by the
Intergalactic Medium.  The simulations were performed with a modified
version of {\small GADGET-3} in which we have incorporated the effect
of neutrinos by using a particle and a grid based method.

We have investigated a wide range of numerical issues relevant  to simulating 
the effect of the free-streaming of neutrinos on the matter
distribution. We find that with a particle based
implementation the spatial distribution of the fast moving neutrinos
suffers significantly from Poisson noise. On scales of interest for
the use of \lya forest data the corresponding errors are, however,
still smaller than the measurement errors of the flux power
spectrum. The less CPU and memory demanding grid based implementation
of neutrinos on the other hand does not suffer from Poisson noise but
results in errors in the matter/flux power spectrum due to the assumption
of linear theory for the growth of perturbations in the neutrino
density. The error in the flux power spectrum in this case is as large
as 4\%, larger or comparable to the measurement errors.
At linear scales $k<0.6 h/$Mpc simulations with the two different
neutrino implementations agree at the 2\% level at $z=0$ and at the
1\% level at $z=3$.  The impact of  other numerical effects
investigated (starting redshift, velocities of the neutrino particles
in the initial conditions) is also smaller or comparable to the
statistical errors of the SDSS flux power spectrum.

By extracting a set of realistic mock quasar  spectra, we
quantify the effect of neutrinos on the flux probability distribution
function and flux power spectrum.  The free-streaming of
neutrinos results in a (non-linear) scale-dependent suppression of
the power spectrum of the total matter distribution at scales probed by
Lyman-$\alpha$ forest data which is larger than the linear theory
prediction by about 25~\% and strongly redshift dependent. The
differences in the matter power spectra translate into a $\sim2.5\%$
($5\%$) difference in the flux power spectrum for neutrino masses with
$\Sigma m_{\nu} = 0.3$ eV (0.6 eV).

We have performed a detailed comparison between simulations including
neutrinos and simulations without neutrinos with a reduced overall
amplitude of the matter power spectrum in order to disentangle as much
as possible the effect of the free-streaming of neutrinos on the shape
of the flux power spectrum and its evolution from the overall
suppression of the power spectrum due to the free-streaming. The latter
is responsible for the well known degeneracy between the (upper limit
for the) mass of neutrinos and the value of $\sigma_8$.  Breaking this degeneracy
is important for a reliable assessment of the robustness of the upper
limits on neutrino masses from \lya forest data as the lowest upper
limits are based on combining measurements of the matter power
spectrum on different scales with very different methods.

We find that the differences in the flux power spectrum and the flux
probability distribution function between simulations including
neutrinos and simulations without neutrinos with a reduced overall
amplitude of the matter power spectrum are small but
noticeable. Motivated by our findings, we then investigated whether
the present SDSS data set {\em alone} can give constraints on the
neutrino masses. We have explored the multi-dimensional
likelihood space using the flux derivative method proposed by
\cite{vh06}. We found a conservative upper limit of \smnu = 0.9 eV at the
$2\sigma$ level, obtained from SDSS quasar spectra alone, which is comparable to limits obtained with other
probes of large scale structure. This limit is of course much weaker
than published constraints obtained by combining \lya forest data with
information from large scales, because the latter leverages the
different r.m.s. value for the amplitude of the matter power spectrum
suggested by small-scale and large-scale observables and turns this
into a tight constraint for the absolute neutrino masses.  The robustness of these recently published relatively low 
upper limits depends, however, strongly on the somewhat questionable assumption
that there are no systematic offsets between the measurements obtained
on small and large scales with these very different methods which are
not yet fully understood or not correctly taken into account in the
error analysis.

We have demonstrated here that a quantitative investigation of the
effect of the free-streaming of neutrinos on the non-linear matter
distribution as probed by the IGM structures can be efficiently
performed with numerical hydrodynamical simulations. Reaching an accuracy
below the one percent level at scales relevant for \lya forest or weak
lensing data will still be challenging but should be doable and will
be an important step in turning the exciting prospect of an actual
measurement of neutrino masses into reality.

\acknowledgments {We thank the referee for a useful and constructive
  report. MV is partly supported by ASI/AAE, INFN-PD51 and a PRIN by
  MIUR. Numerical computations were performed using the COSMOS
  Supercomputer in Cambridge (UK), which is sponsored by SGI, Intel,
  HEFCE and the Darwin Supercomputer of the University of Cambridge
  High Performance Computing Service (http://www.hpc.cam.ac.uk/),
  provided by Dell Inc. using Strategic Research Infrastructure
  Funding from the Higher Education Funding Council for England. Part
  of the analysis has been performed at CINECA with a Key project on
  the intergalactic medium obtained through a CINECA/INAF grant.}

\bibliographystyle{JHEP.bst}

\bibliography{nu.bib}

\end{document}